%% file: paper.tex
\documentclass[structabstract]{aa}  
\usepackage{graphicx}
\usepackage{subfigure}
\usepackage{amsmath}
\usepackage[english]{babel}
\usepackage{natbib}
\usepackage{longtable}
\usepackage{lscape}
\bibpunct{(}{)}{;}{a}{}{,}
\usepackage{txfonts}
\newcommand{\vect}[1]{\ensuremath{\mathbf{#1}}}

\newcommand{\bv}{\mathbf{v}}
\newcommand{\bb}{\mathbf{B}}
\newcommand{\mrho}{\overline{\varrho}}
\newcommand{\mT}{\overline{T}}\newcommand{\mP}{\overline{P}}
\newcommand*{\dif}{\mathop{}\!\mathrm{d}}
\newcommand{\Prandtl}{\ensuremath{Pr}}
\newcommand{\Pm}{\ensuremath{Pm}}

\newcommand{\Ro}{\ensuremath{Ro}}
\newcommand{\Rol}{\ensuremath{Ro_\ell}}
\newcommand{\lc}{\ensuremath{\ell_c}}
\newcommand{\Roz}{\ensuremath{Ro_z}}
\newcommand{\Ra}{\ensuremath{Ra}}
\newcommand{\Rac}{\ensuremath{Ra_\text{c}}}

\newcommand{\fdip}{\ensuremath{f_\text{dip}}}
\newcommand{\fdipAX}{\ensuremath{{f_\text{dip}}_\text{ax}}}
\newcommand{\Lo}{\ensuremath{Lo}}
\newcommand{\Elsasser}{\ensuremath{\Lambda}}
\newcommand{\fohm}{\ensuremath{f_\text{ohm}}}
\newcommand{\Lof}{\ensuremath{Lo/\sqrt{f_{\text{ohm}}}}}
\newcommand{\Nrho}{\ensuremath{N_\varrho}}
\newcommand{\chir}{\ensuremath{\chi_{rel}}}
\newcommand{\E}{\ensuremath{\text{E}}}
\newcommand{\Ek}{\ensuremath{E_\mathrm{k}}}
\newcommand{\Em}{\ensuremath{E_\mathrm{m}}}
\newcommand{\ro}{\ensuremath{r_{\!o}}}
\newcommand{\aspectratio}{\ensuremath{\chi}}
\newcommand{\mm}{\ensuremath{\mathrm{m}}}
\newcommand{\dd}{\ensuremath{\mathrm{d}}}
\newlength{\mafig}
\newlength{\figsl}
\begin{document}
   \title{Influence of the mass distribution on the magnetic field
   topology}

   \author{R. Raynaud\inst{\ref{i1},\ref{i2}} 
     \and L. Petitdemange\inst{\ref{i1},\ref{i2}} 
     \and E. Dormy\inst{\ref{i1},\ref{i3}}
   } 

   \institute{ MAG (ENS / IPGP), LRA, D\'epartement de
     Physique, \'Ecole normale sup\'erieure, 24 rue Lhomond, 75252
     Paris Cedex 05, France \label{i1} \and LERMA, CNRS UMR 8112,
     Paris, France \label{i2} \and IPGP, CNRS UMR 7154, Paris,
     France \\ 
     \email{raphael.raynaud@ens.fr,
     ludovic@lra.ens.fr, 
     dormy@phys.ens.fr}\label{i3}
     }

   \date{Received 28 March 2014 / Accepted 16 June 2014}

 
  \abstract 
  {Three-dimensional spherical dynamo simulations carried out within
  the framework of the anelastic approximation have revealed that the
  established distinction between dipolar and multipolar dynamos tends
  to be less clear than it was in Boussinesq studies. This result was
  first interpreted as a direct consequence of the existence of a
  larger number of models with a high equatorial dipole contribution,
  together with an intermediate dipole field strength. However, this
  finding has not been clearly related to specific changes
  that would have been introduced by the use of the anelastic
  approximation.}
  {In this paper, we primarily focus on the effects of choosing of a
    different mass distribution.  Indeed, it is likely to have as
    large consequences as would taking a stratified reference state into
    account, especially when comparing our results to previous
    Boussinesq studies. }
  {Our investigation is based on a systematic parameter study of 
   weakly stratified anelastic dynamo models.}
  {We show that the tendencies highlighted in previous anelastic
  dynamo simulations are already present in the Boussinesq limit. Thus
  they cannot be systematically related to anelastic
  effects. Actually, a central mass distribution can result in changes
  in the magnetic field topology that are mainly due to the
  concentration of convective cells close to the inner sphere.}
   {} 

   \maketitle
%
\section{Introduction}

Dynamo action, i.e. the self-amplification of a magnetic field by the
flow of an electrically conducting fluid, is considered to be the main
mechanism for generating magnetic fields in the universe for a variety
of systems, including planets, stars, and
galaxies~\citep[e.g.][]{book_dormy}. Dynamo action is an instability
by which a conducting fluid transfers part of its kinetic energy to
magnetic energy.  Because of the difficulty simulating turbulent fluid
motions, one must resort to some approximations to model the fluid
flow, whose convective motions are assumed to be driven by the
temperature difference between a hot inner core and a cooler outer
surface. A strong simplification can be achieved when applying the
Boussinesq approximation \citep{boussinesq1903}, which performs well
in so far as variations in pressure scarcely affect the density of the
fluid.  However, in essence, this approximation will not be adequate
for describing convection in highly stratified systems, such as
stars or gas giants. A common approach to overcoming this difficulty
is then to use the anelastic approximation, which allows for a
reference density profile while filtering out sound waves for faster
numerical integration.  This approximation was first developed to
study atmospheric convection \citep{ogura62,gough69}. It has then been
used to model convection in the Earth core or in stars and is found in
the literature under slightly different formulations
\citep{gilman81,braginsky95,lantz99,anufriev2005,
  berkoff2010,jones11,alboussiere13}. Nevertheless, the starting point
in the anelastic approximation is always to consider convection as a
perturbation of a stratified reference state that is assumed to be
close to adiabatic.

Observations of low mass stars have revealed very different magnetic
field topologies from small scale fields to large scale dipolar
fields \citep{donati09,morinD2010}, and highlight possible
correlations between differential rotation and magnetic field
topologies \citep{reinhold2013}.  Boussinesq models partly reproduce
this diversity \citep{busse06,morin11,sasaki2011,schrinner12}.
Moreover, the dichotomy between dipolar and ``non-dipolar'' (or
multipolar) dynamos seems to hold for anelastic models
\citep{gastine12,yadav13}.  However, we show in \cite{schrinner2014}
that this distinction may somewhat be less clear than it was with
Boussinesq models. Indeed, in a systematic parameter study we found a
large number of models with both a high equatorial dipole contribution
and an intermediate dipole field strength. Only a few examples of
equatorial dipoles have been reported from Boussinesq spherical dynamo
simulations \citep{aubert04, gissinger12}.
At the same time, observations have shown that for some planets, such
as Uranus or Neptune, the dipole axis can make an angle up to $\pi/2$
with respect to the rotation axis, owing to a significant contribution
from the equatorial dipole \citep{jones2011a}.

In this paper, we aim to clarify the reasons likely for the emergence
of an equatorial dipole contribution when measuring the dipole field
strength at the surface of numerical models. Since our approach
closely follows previous methodology for studying the link with
Boussinesq results, we decided to focus in more detail on one
important change that comes with the anelastic approximation as
formulated in \cite{jones11}, assuming that all mass is concentrated
inside the inner sphere to determine the gravity profile. In contrast,
as proposed by the Boussinesq dynamo benchmark \cite{christensen01},
it was common for geodynamo studies to assume that the density is
homogeneously distributed. This leads to different gravity profiles,
the first being proportional to $1/r^2$, whereas the second is
proportional to $r$.  According to
\cite{duarte2013}, \cite{gastine12} show that both gravity profiles
lead to very similar results. Contrary to this statement, we show that
the choice of the gravity profile may have strong consequences on the
dynamo-generated field topology.  We briefly recall the anelastic
equations in Section~\ref{s_eq} before presenting our results in
Section~\ref{s_res}. In Appendix~\ref{s_scaling}, we give the fit
coefficients obtained for the scalings of the magnetic and velocity
fields and the convective heat flux. A summary of the numerical
simulations carried out is given in Appendix~\ref{s_num}.

\section{Governing equations}
\label{s_eq}
\subsection{The non-dimensional anelastic equations}
We rely on the LBR-formulation, named after \cite{lantz99} and
\cite{braginsky95}, as it is used in the dynamo benchmarks proposed by
\cite{jones11}. It guarantees the energy conservation, unlike
other formulations \citep[see][]{brown12}. A more detailed
presentation of the equations can be found in our preceding paper
\cite{schrinner2014}.

Let us consider a spherical shell of width~$d$ and aspect
ratio~$\aspectratio$, rotating about the $z$ axis at angular
velocity~$\vect{\Omega}$ and filled with a perfect, electrically
conducting gas with kinematic viscosity~\(\nu\), thermal
diffusivity~\(\kappa\), specific heat~$c_p$, and magnetic
diffusivity~\(\eta \) (all assumed to be constant).  In contrast to
the usual Boussinesq framework, convection is driven by an imposed
entropy difference~$\Delta s$ between the inner and the outer
boundaries, and the gravity is given by
\(\mathbf{g}=-GM\mathbf{\hat{r}}/r^2\) where $G$ is the gravitational
constant and $M$ the central mass, assuming that the bulk of the mass
is concentrated inside the inner sphere.

The reference state is given by the polytropic equilibrium solution of
the anelastic system
\begin{equation}
\mP=P_c\,w^{n+1},\quad\mrho=\varrho_c\,w^n,\quad \mT=T_c\,w,\quad 
w=c_0+\frac{c_1 d}{r}, 
\label{ref_state}
\end{equation}
\begin{equation}
c_0=\frac{2w_0-\aspectratio-1}{1-\aspectratio},\quad
c_1=\frac{(1+\aspectratio)(1-w_o)}{(1-\aspectratio)^2},
\label{c0_c1}
\end{equation}
with 
\begin{equation}
w_0=\frac{\aspectratio+1}{\aspectratio\exp(N_\varrho/n)+1},\quad
w_i=\frac{1+\aspectratio-w_o}{\aspectratio}\,.
\label{w1_w0}
\end{equation}
In the above expressions, $n$ is the polytropic index and \Nrho{} the
number of density scale heights, defined by
\(N_\varrho=\ln{(\varrho_i/\varrho_o)}\), where \(\varrho_i\) and
\(\varrho_o\) denote the reference state density at the inner and
outer boundaries, respectively. The values $P_c$, $\varrho_c$, and
$T_c$ are the reference-state density, pressure, and temperature midway
between the inner and outer boundaries, and serve as units for these
variables.

We adopt the same non-dimensional form as \cite{jones11}: length is
scaled by the shell width~$d$, time by the magnetic diffusion time
~$d^2/\eta$, and entropy by the imposed entropy difference~$\Delta
s$.  The magnetic field is measured in units of
\(\sqrt{\Omega\varrho_c\mu\eta}\) where \(\mu\) is the magnetic
permeability.  Then, the equations governing the system are
\begin{align}
  \begin{split}\label{mhd1}
    \frac{\partial\bv}{\partial t} + \bv\cdot\nabla\bv &=
    Pm\,
    \bigg[-\frac{1}{E}\nabla\frac{P'}{w^n}
        +\frac{Pm}{Pr}Ra\frac{s}{r^2} \mathbf{\hat{r}}
        -\frac{2}{E}\,\mathbf{\hat{z}} \times\bv \\ 
        &\quad
        + \mathbf{F}_\nu+\frac{1}{E\,w^n}(\nabla\times\bb)\times\bb
        \bigg]\,,
  \end{split} \\
  \frac{\partial\bb}{\partial t} &= 
  \nabla\times(\bv\times\bb)+\nabla^2\bb \,,\label{mhd2}\\
  \begin{split}\label{mhd3}
    \frac{\partial s}{\partial t}+\bv\cdot\nabla s &= 
    w^{-n-1}\frac{Pm}{Pr}\nabla\cdot\left(w^{n+1}\,\nabla s\right) \\ 
    &\quad + \frac{Di}{w}\left[E^{-1}w^{-n}(\nabla\times \bb)^2+Q_\nu\right]\,,
  \end{split}\\
  \nabla\cdot \left(w^n\bv \right)  &= 0\,,\label{mhd4}\\ 
  \nabla\cdot\bb &=  0\,.\label{mhd5}
\end{align}
The viscous force $\vect{F}_\nu$ in \eqref{mhd1} is given by
\(\mathbf{F}_\nu=w^{-n}\nabla\mathbf{S}\), where $\vect{S}$ is the rate of
strain tensor
\begin{equation}
  S_{ij}=2w^n\left(e_{ij}-\frac{1}{3}\delta_{ij}\nabla\cdot \bv\right),\quad
  e_{ij}=\frac{1}{2}\left(\frac{\partial v_i}{\partial x_j}+\frac{\partial
    v_j}{\partial x_i}\right) \, .
\end{equation}
Moreover, the expressions of the dissipation parameter~\(Di\) and the
viscous heating~\(Q_\nu\) in (\ref{mhd3}) are
\begin{equation}
Di=\frac{c_1Pr}{Pm Ra} \, ,
\end{equation} 
and
\begin{equation}
Q_\nu=2\left[e_{ij}e_{ij}-\frac{1}{3}(\nabla\cdot\bv)^2\right] \, .
\end{equation}
The boundary conditions are the following. We impose stress free
boundary conditions for the velocity field at both the inner and the
outer sphere, the magnetic field matches a potential field inside and
outside the fluid shell, and the entropy is fixed at the inner and
outer boundaries.

The system of equations (\ref{mhd1})-(\ref{mhd5}) involves seven
control parameters, namely the Rayleigh number~\(Ra=GMd\Delta s /
(\nu\kappa c_p)\), the Ekman number~\(E =\nu / (\Omega d^2) \), the
Prandtl number~\(Pr = \nu / \kappa \), and the magnetic Prandtl
number~\(Pm = \nu / \eta\), together with the aspect
ratio~\aspectratio{}, the polytropic index~\(n\), and the number of
density scale heights~\(N_\varrho\) that define the reference state.
In this study, we restrict our investigation of the parameter space
keeping $\E=10^{-4}$, $\Prandtl=1$, $\aspectratio=0.35$, and $n=2$ for
all simulations. Furthermore, to differentiate the effects related to
the change in gravity profile from those related to the anelastic
approximation, we decided to perform low $\Nrho$ simulations so that
we can assume that stratification no longer influences the dynamo
process. In practice, we chose $\Nrho=0.1$, which means that
the density contrast between the inner and outer spheres is only
1.1, and the simulations are thus very close to the Boussinesq limit.
To further ensure the lack of stratification effects, we also checked
in a few cases that the results do not differ from purely Boussinesq
simulations.


We have integrated our system at least on one magnetic diffusion time
with the anelastic version of \textsc{parody} (\cite{dormy98} and
further developments), which reproduces the anelastic dynamo
benchmarks \citep[see][]{schrinner2014}.  The vector fields are
transformed into scalars using the poloidal-toroidal
decomposition. The equations are then discretized in the radial
direction with a finite-difference scheme; on each concentric sphere,
variables are expanded using a spherical harmonic basis. The
coefficients of the expansion are identified with their degree~$\ell$
and order~$m$. Typical resolutions use from 256 to 288 points in the
radial direction and a spectral decomposition truncated at $80 \le
\ell_\text{max} \sim m_\text{max} \le 116$.


\subsection{Diagnostic parameters}
The quantities used to analyse our simulations first rely on the
kinetic and magnetic energy densities \Ek{} and \Em{},
\begin{equation}
\Ek = \frac{1}{2\, V}\int_V\,w^n\bv^2 \dif v 
\quad \text{and} \quad
\Em = \frac{1}{2\, V}\frac{Pm}{E}\int_V\,\bb^2 \dif v
\,.
\label{emag}
\end{equation}
We define the corresponding Rossby number $Ro=\sqrt{2 \Ek} E/Pm$ and
Lorentz number $\Lo=\sqrt{2 \Em}E/Pm$. The latter is a non-dimensional
measure of the magnetic field amplitude, while the former is a
non-dimensional measure of the velocity field amplitude. A measure of
the mean zonal flow is the zonal Rossby number \Roz{}, whose
definition is based on the averaged toroidal axisymmetric kinetic
energy density. To distinguish between dipolar and multipolar dynamo
regimes, we know from Boussinesq results that it is useful to measure
the balance between inertia and Coriolis force, which can be
approximated in terms of a local Rossby number $ \Rol = Ro_c\,\lc/\pi
$, which depends on the characteristic length scale of the flow rather
than on the shell thickness
\citep{christensen06,olson06,schrinner12}. Again, we emphasize
that our definition of the local Rossby number tries to avoid any
dependence on the mean zonal flow and thus differs from the original
definition introduced by \cite{christensen06}\citep[see][App.~A for a
  discussion]{schrinner12}. Our typical convective length scale is
based on the mean harmonic degree~\lc{} of the velocity component
$\vect{v}_c$ from which the mean zonal flow has been subtracted,
\begin{equation}
  \lc = \sum_\ell \ell\frac{< w^n\, (\bv_c)_\ell\cdot(\bv_c)_\ell>}
      {< w^n \, \bv_c\cdot\bv_c>}
      \,.
\end{equation}
where the brackets denote an average of time and radii. Consistently,
the contribution of the mean zonal flow is removed for calculating
$\Ro_c$. 

The dipolarity of the magnetic field is characterized by the relative
dipole field strength, \fdip{}, originally defined as the time-average
ratio on the outer shell boundary $S_{\!o}$ of the dipole field
strength to the total field strength,
\begin{equation}
  \fdip = \left\langle
  \sqrt{
  \frac{\int_{S_{\!o}} {\vect{B}^2}_{\,\ell=1}^{\,m=\left \{0,1\right \}} 
    \, \sin \theta \dif \theta \dif \phi }{\int_{S_{\!o}} \vect{B}^2 
    \, \sin \theta \dif \theta \dif \phi }
  }
  \right\rangle_t
  \,.
\end{equation}
We also define a relative \emph{axial} dipole field strength filtering
out non-axisymmetric contributions
\begin{equation}
  \fdipAX = \left\langle
  \sqrt{
  \frac{\int_{S_{\!o}} {\vect{B}^2}_{\,\ell=1}^{\,m=0} 
    \, \sin \theta \dif \theta \dif \phi }{\int_{S_{\!o}} \vect{B}^2
    \, \sin \theta \dif \theta \dif \phi } 
  }
  \right\rangle_t
  \,.\label{e_fdipAX}
\end{equation}
This definition of \fdipAX{} is similar to the relative dipole field
strength used by \cite{gastine12}, except for the square root, which
explains the lower values for the dipolarity found in \cite{gastine12}.

To further characterize the topology of the magnetic field, we
introduce a time-averaged measure of the departure from a pure
equatorial dipole solution
\begin{equation}
  \theta = \frac{2}{\pi} \left\langle \sqrt{\left(\Theta(t) - \frac{\pi}{2}
    \right)^2}\right\rangle_t
  \,,\label{e_theta}
\end{equation}
where $\Theta$ is the tilt angle of the dipole axis.
A low value of $\theta$ indicates that the tilt angle of the dipole
fluctuates close to $\Theta = \pi/2$.

\section{Results}
\label{s_res}

\subsection{Bifurcations between dynamo branches}
\begin{figure}[t]
     \centering
    (a) 
    \includegraphics[width=\mafig]{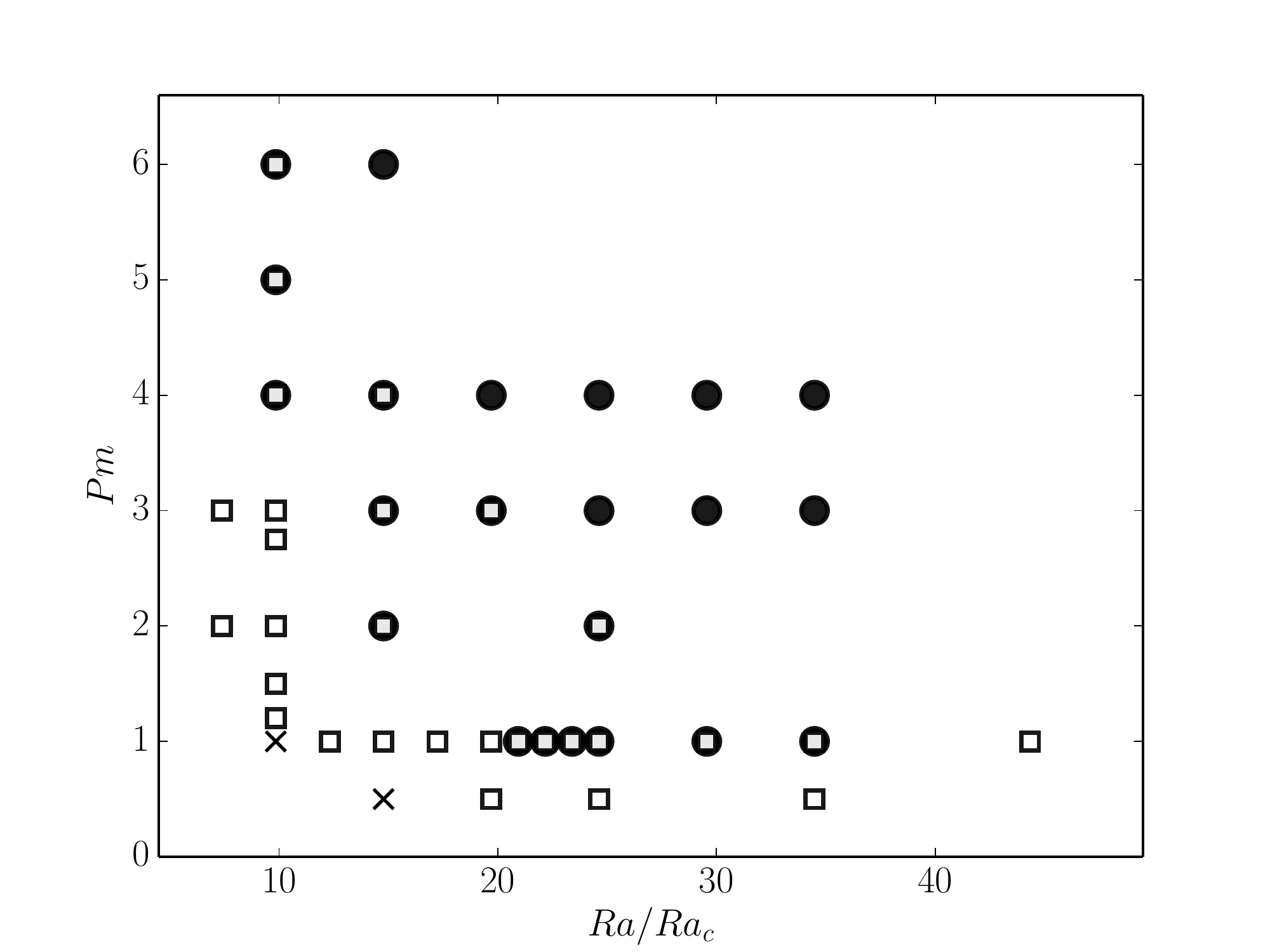}
    (b) 
    \includegraphics[width=\mafig]{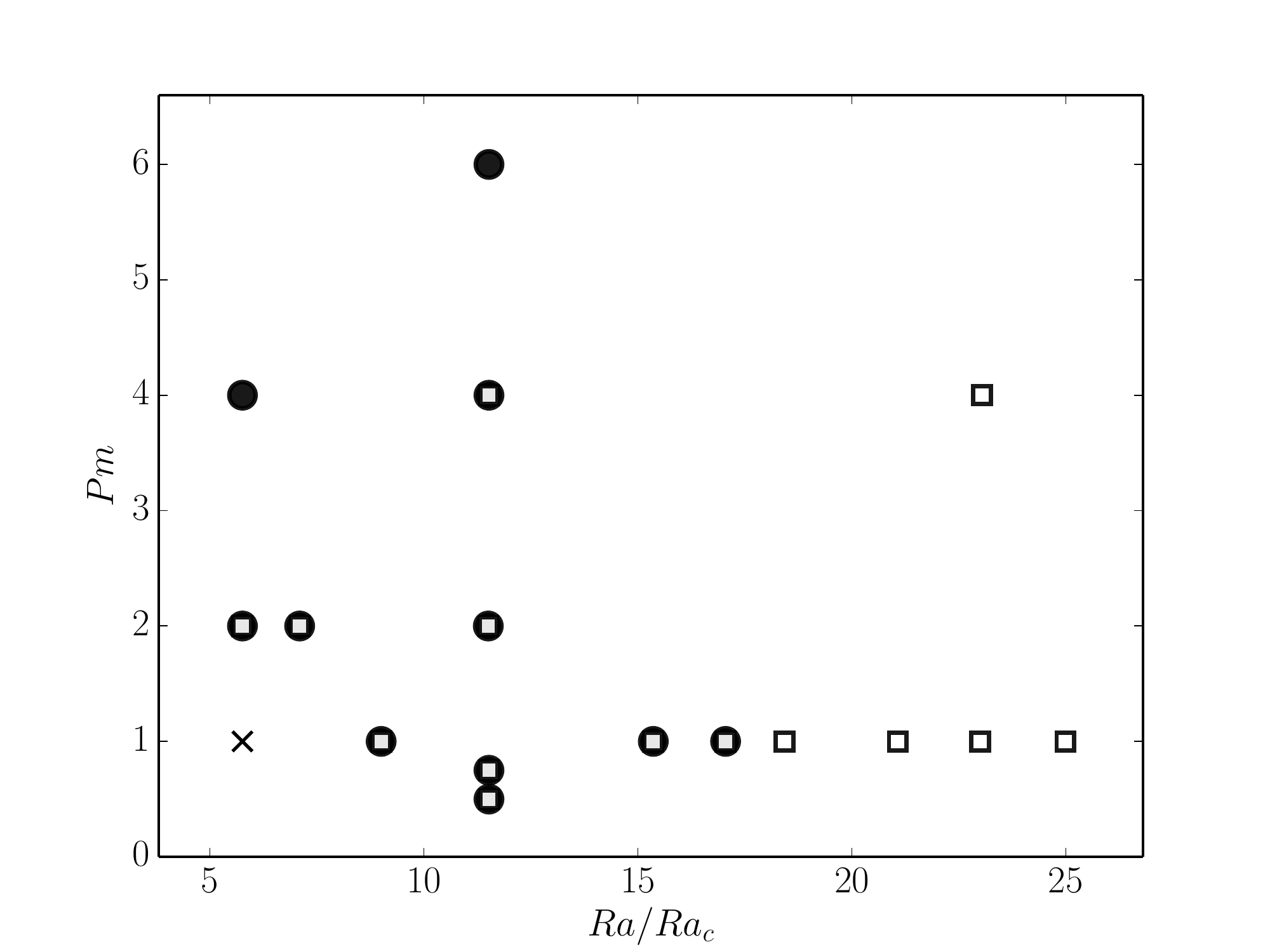} 
    
    \caption{Dipolar (black circles) and multipolar (white squares)
    dynamos as a function of $\Ra/\Rac$ and $\Pm$,
    for a central mass (a) and a uniform mass distribution
    (b). Crosses indicate the absence of a self-sustained
    dynamo. }\label{f_RA_PM_01}
\end{figure}
In our simulations we recover the two distinct dynamo regimes
observed with both Boussinesq \citep{kutzner02,christensen06,
schrinner12, yadav12} and anelastic
models \citep{gastine12,schrinner2014}.  These are characterized by
different magnetic field configurations: dipolar dynamos are dominated
by a strong axial dipole component, whereas ``non-dipolar'' dynamos
usually present a more complex geometry with higher spatial and
temporal variability.  The branches are easily identified by
continuing simulations performed with other parameters, for which the
dipolar/multipolar characteristic was previously established.
Figure~\ref{f_RA_PM_01}(a) shows the regime diagram we obtained, as a
function of the Rayleigh and magnetic Prandtl numbers.  For
comparison, we use the data from \cite{schrinner12} and show in
Fig.~\ref{f_RA_PM_01}(b) the same regime diagram obtained for
Boussinesq models with a uniform mass distribution. For $Pm=1$, the
transition from the dipolar to the multipolar branch can be triggered
by an increase in \Ra{}. In that case, the transition is due to the
increasing role of inertia as revealed by \Rol{}. Alternatively, the
transition from multipolar to dipolar dynamo can be triggered by
increasing \Pm{}. Then, the multipolar branch is lost when the
saturated amplitude of the mean zonal flow becomes too small to
prevent the growth of the dipolar solution \citep[see][]{schrinner12}.
It is worth noting that the two branches overlap for a restricted
parameter range for which dipolar and multipolar dynamos may
coexist. In that case, the observed solution strongly depends of the
initial magnetic field, so we tested both weak and strong field
initial conditions for all our models to delimit the extent of the
bi-stable zone with greater accuracy. Actually, multipolar dynamos are
favoured by the stronger zonal wind that may develop with stress-free
boundary conditions, allowing for this hysteretic
transition \citep{schrinner12}. Finally, we see that the dynamo
threshold is lower for multipolar models, which allows the multipolar
branch to extend below the dipolar branch at low Rayleigh and
magnetic Prandtl numbers. We see in Fig.~\ref{f_RA_PM_01}(b) that this
is different from Boussinesq models with a uniform mass distribution.

\begin{figure}[htbp]
 
 \centering 
 (a)
 \includegraphics[width=\mafig]{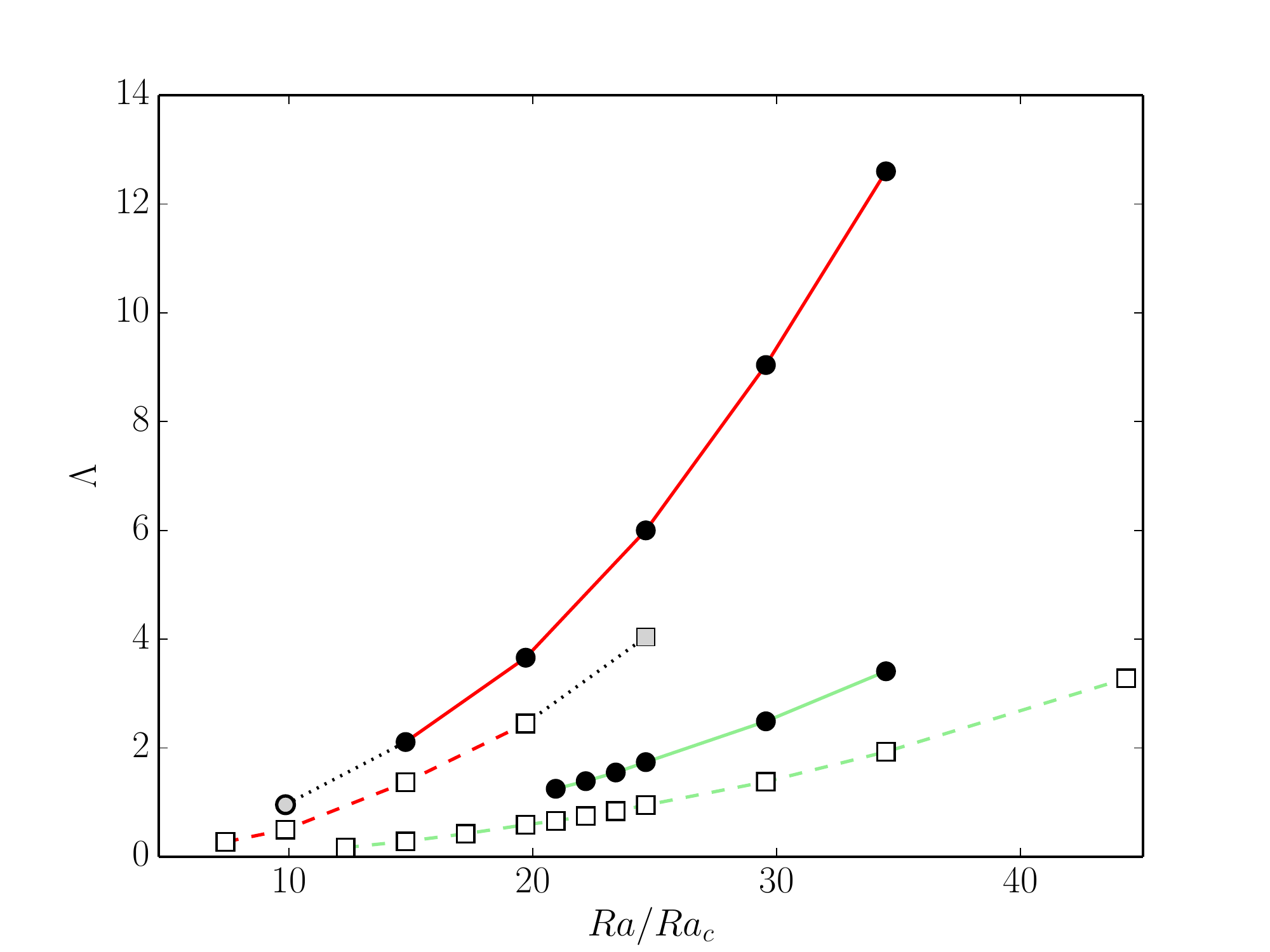}
 (b)
 \includegraphics[width=\mafig]{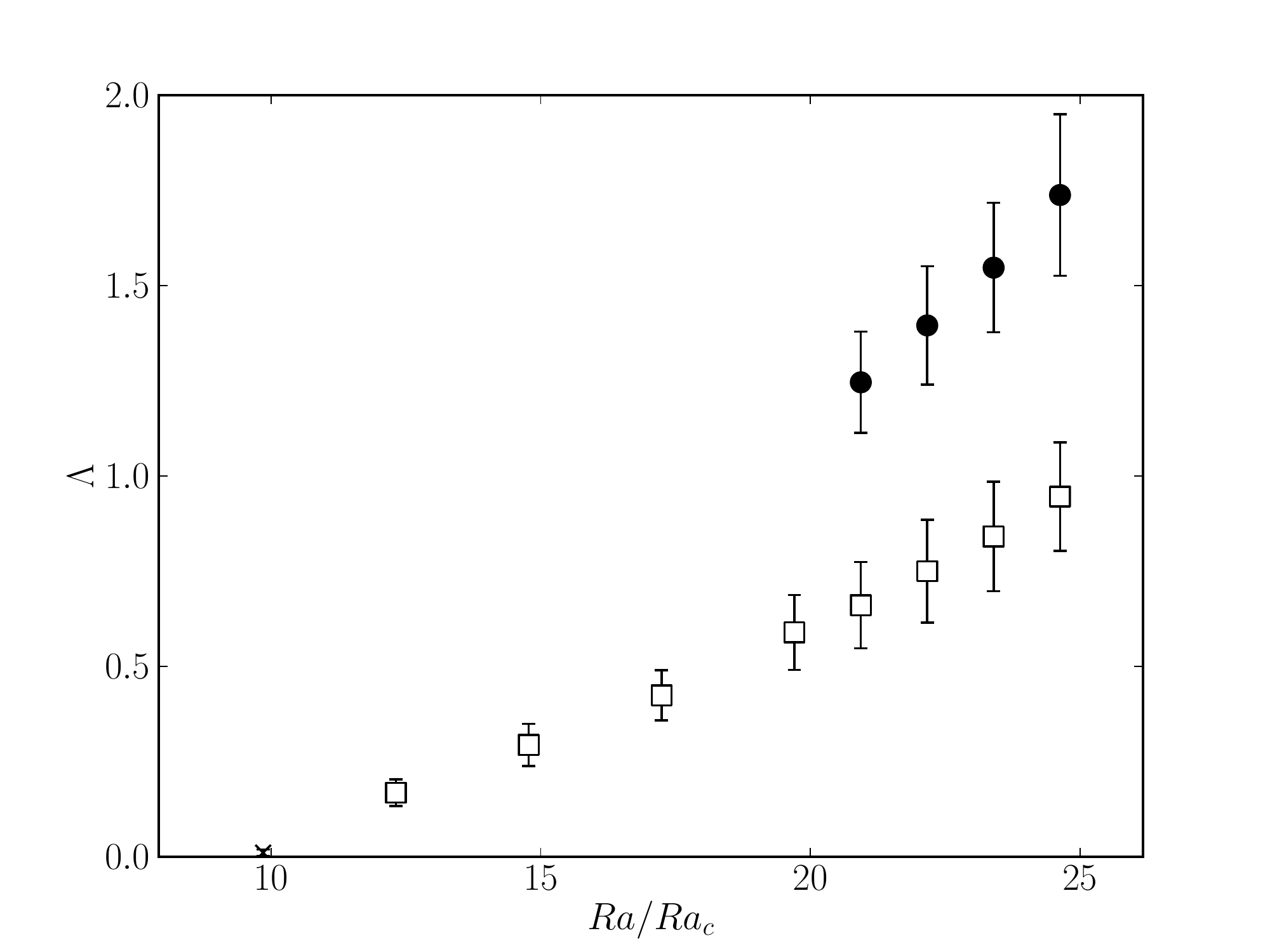}

 \caption{(a): Elsasser number \Elsasser{} as a function of $\Ra/\Rac$, for
 $\Pm=1$ (green) and $\Pm=3$ (red).  The meaning of the symbol shapes
 is defined in the caption of Fig.~\ref{f_RA_PM_01}. A grey marker
 indicates that the solution loses its stability. (b):
 Detail of the bifurcation close to the dynamo threshold for
 $\Pm=1$. Error bars represent the standard deviations.  }\label{f_b}
\end{figure}

To investigate the different transitions between the different dynamo
branches, we plot the Elsasser number $\Elsasser =
B^2_\text{rms}/(\Omega \varrho_c \mu \eta) $ in Fig.~\ref{f_b}(a)
(related to the Lorentz number by $\Lambda = \Lo^ 2 \Pm /\E$) as a
function of the distance to the threshold for models at $\Pm=1$ and
$\Pm=3$.
We see in Fig.~\ref{f_b}(b) that the bifurcation for multipolar branch
is supercritical. When decreasing the Rayleigh number, the dipolar
branch loses its stability for $\Ra/\Rac \sim 20$, when the magnetic
field strength becomes too weak. 

\begin{figure}[htbp]
 \centering \includegraphics[width=\mafig]{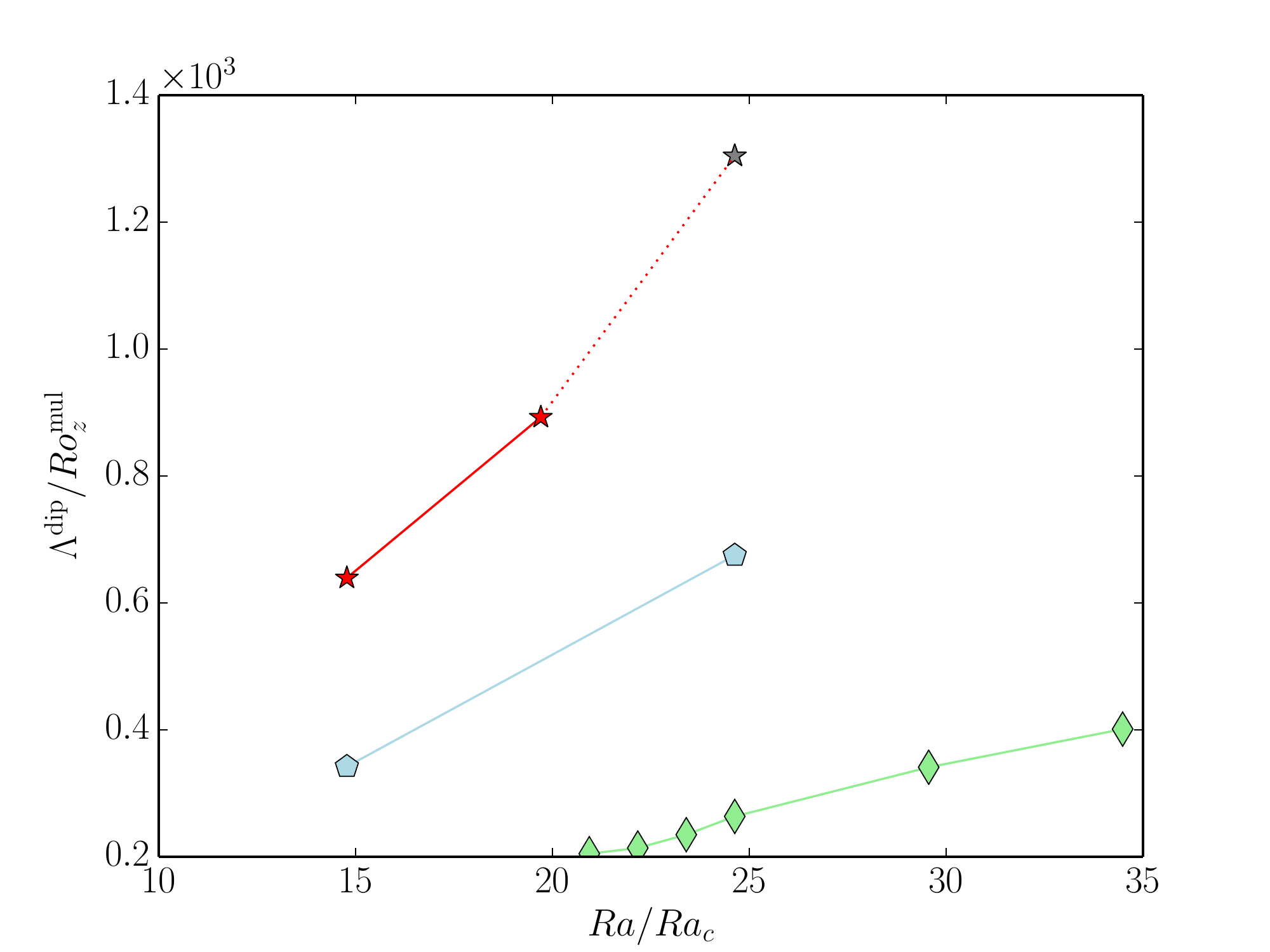}

 \caption{Ratio $\Lambda^\mathrm{dip}/\Roz^\mathrm{mul}$ as a function
 of $\Ra/\Rac$, for $\Pm=1$ (green diamonds), $\Pm=2$ (blue pentagons)
 and $\Pm=3$ (red stars). The point marked with the grey star has been
 computed with the model corresponding to the grey square in
 Fig.~\ref{f_b}.  }\label{f_rapport}
\end{figure}

For higher magnetic Prandtl numbers, the bifurcation of the multipolar
branch still  seems to be supercritical. Interestingly, one notes in
Fig.~\ref{f_b}(a) for $\Pm=3$ that the multipolar branch loses its
stability when increasing the Rayleigh number. A physical explanation
for this behaviour is that the mean zonal flow does not grow fast
enough as the field strength increases, and the dynamo switches to the
dipolar solution. This simple physical scenario can be illustrated by
comparing the variation in the field strength of the dipolar branch,
as measured by $\Lambda^\mathrm{dip}$, and the zonal shear of the
multipolar branch, as measured by $\Roz^\mathrm{mul}$. Indeed, we see
in Fig.~\ref{f_rapport} that the higher the magnetic Prandtl number,
the faster the growth of the ratio between $\Lambda^\mathrm{dip}$ and
$\Roz^\mathrm{mul}$. This explains why the multipolar branch
destabilizes at large forcing for larger \Pm{} ($\Pm=3$, red dashed
line in Fig.~\ref{f_b}(a)), while it remains stable at smaller \Pm{}
($Pm=1$, green dashed line in Fig.~\ref{f_b}(a)).
Because of computational limitations, we were not able to find for
$\Pm>1$ the Rayleigh numbers for which the dipolar branch should
disappear.


\subsection{Equatorial dipole}

\cite{schrinner2014} show that dipolar and multipolar dynamos 
in anelastic simulations were no longer distinguishable from each
other in terms of \fdip{}, contrary to Boussinesq models. This
smoother transition has been attributed to the presence of dynamos
with a high equatorial dipole contribution, which leads to
intermediate values for \fdip{}.
\begin{figure}[htbp]      
  \centering
  (a)
  \includegraphics[width=\mafig]{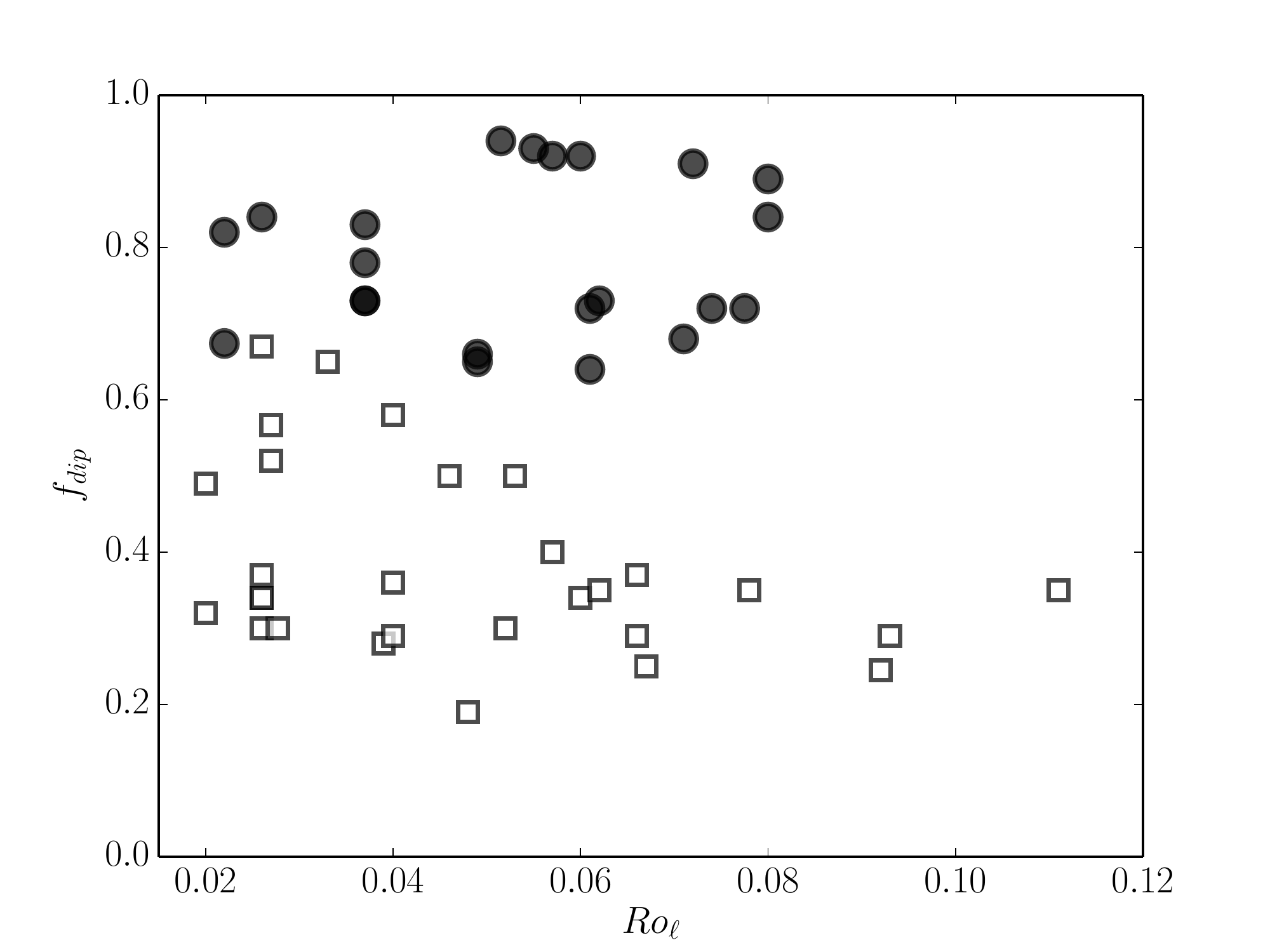}
  (b)
  \includegraphics[width=\mafig]{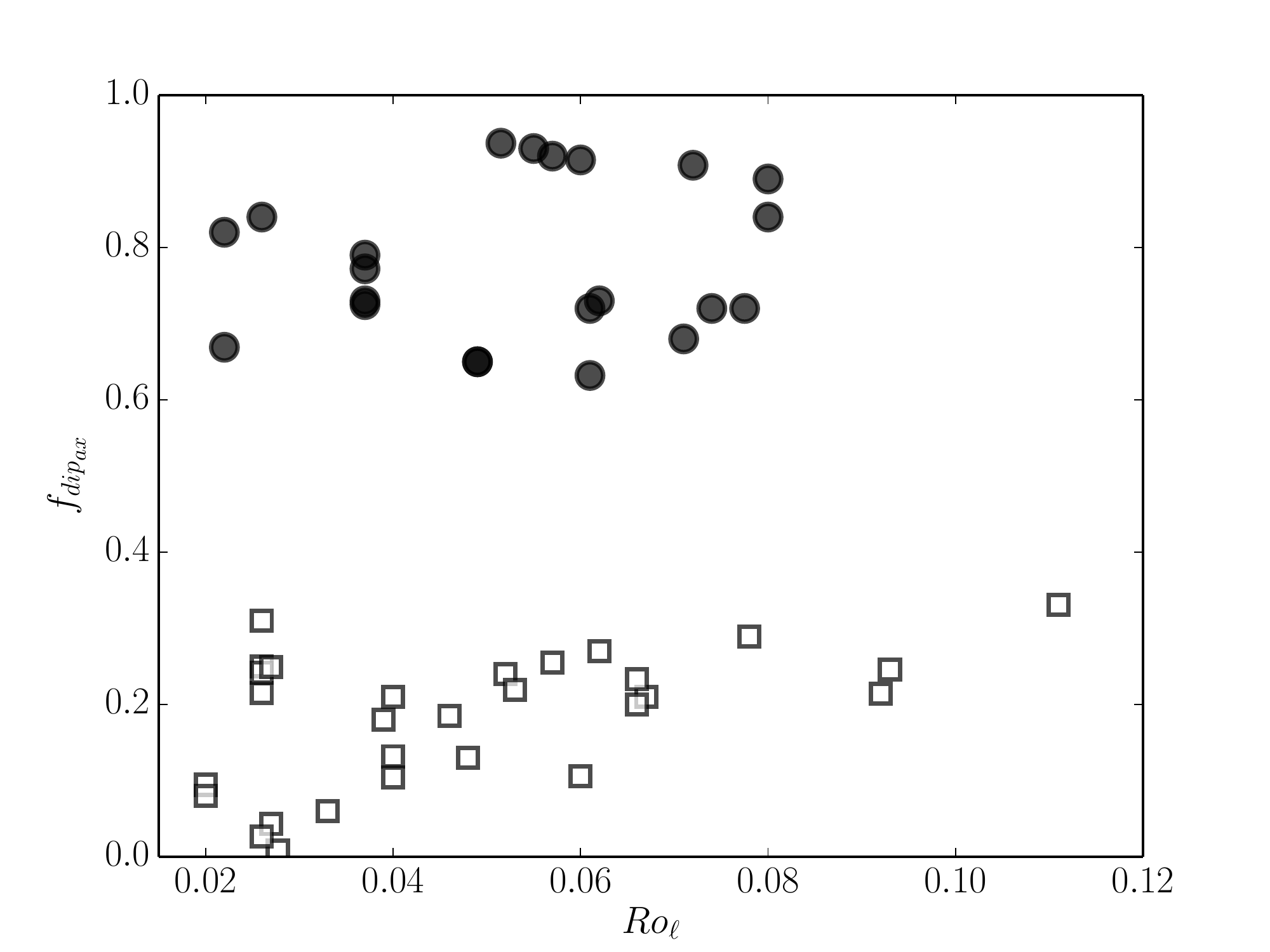}
    
  \caption{(a): The relative dipole field strength \fdip{} versus the
      local Rossby number.  (b): The relative axial dipole field
      strength \fdipAX{} versus the local Rossby number.  The meaning
      of the symbol shapes is defined in the caption of
      Fig.~\ref{f_RA_PM_01}.  }\label{f_fdip}
\end{figure}
However, Fig.~\ref{f_fdip}(a) shows that this tendency already exits
at low $\Nrho$, and thus cannot be accounted for only in terms of
anelastic effects. Furthermore, when the equatorial dipole component
is removed to compute the relative dipole field strength, we recover a
more abrupt transition, as we can see in Fig.~\ref{f_fdip}(b) which
shows the relative axial dipole field strength \fdipAX{}.  Dipolar
dynamos are left unchanged by this new definition, whereas multipolar
dynamos of intermediate dipolarity are no longer observed, which
confirms that the increase in \fdip{} is due to a significant
equatorial dipole component.  The quantity \fdipAX{} therefore
provides a robust criterion to distinguish the dipolar and the
multipolar branches.

\begin{figure}
  \centering
  (a) \includegraphics[width=\mafig]{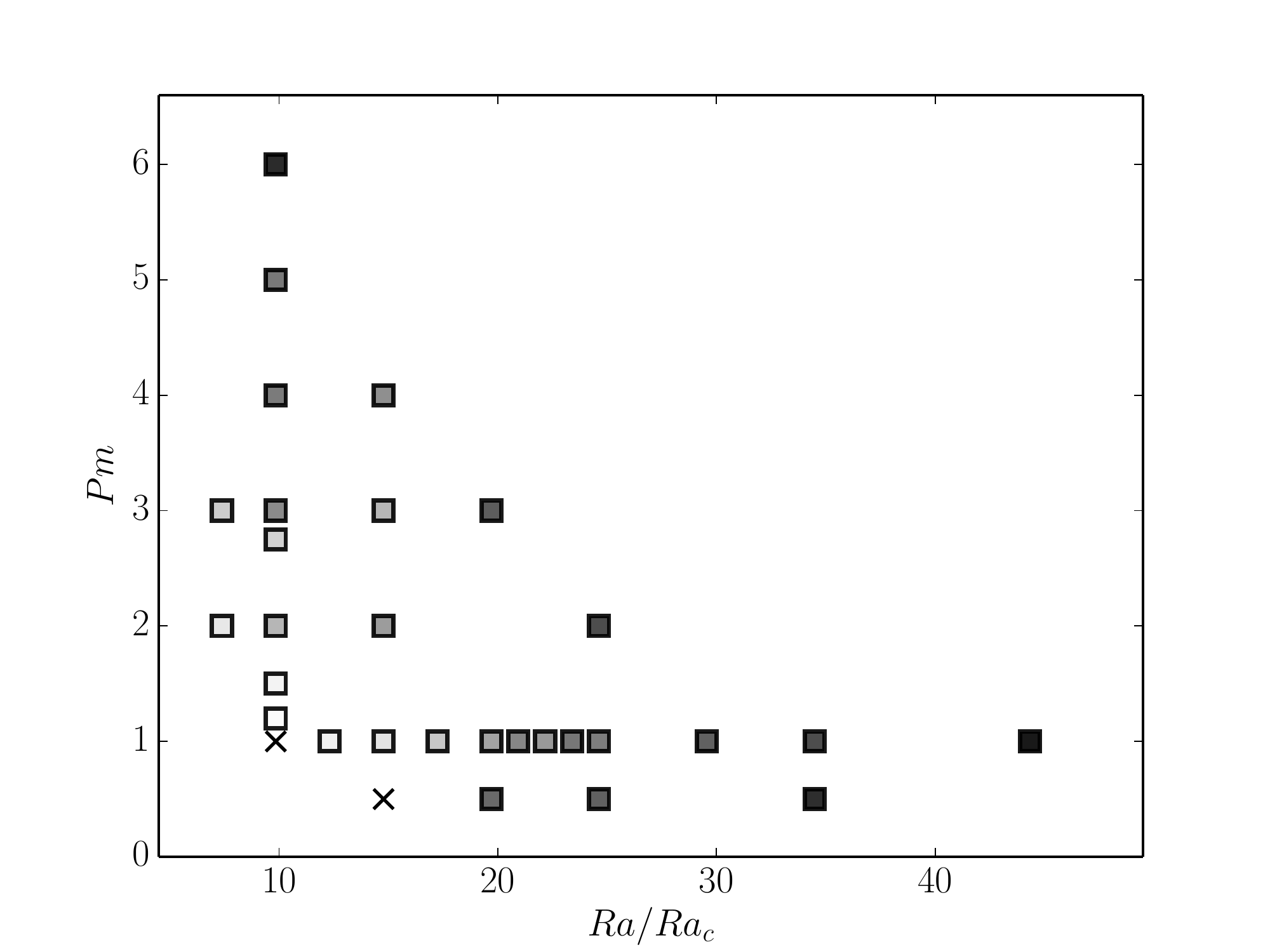} 
  
  (b) \includegraphics[width=\mafig]{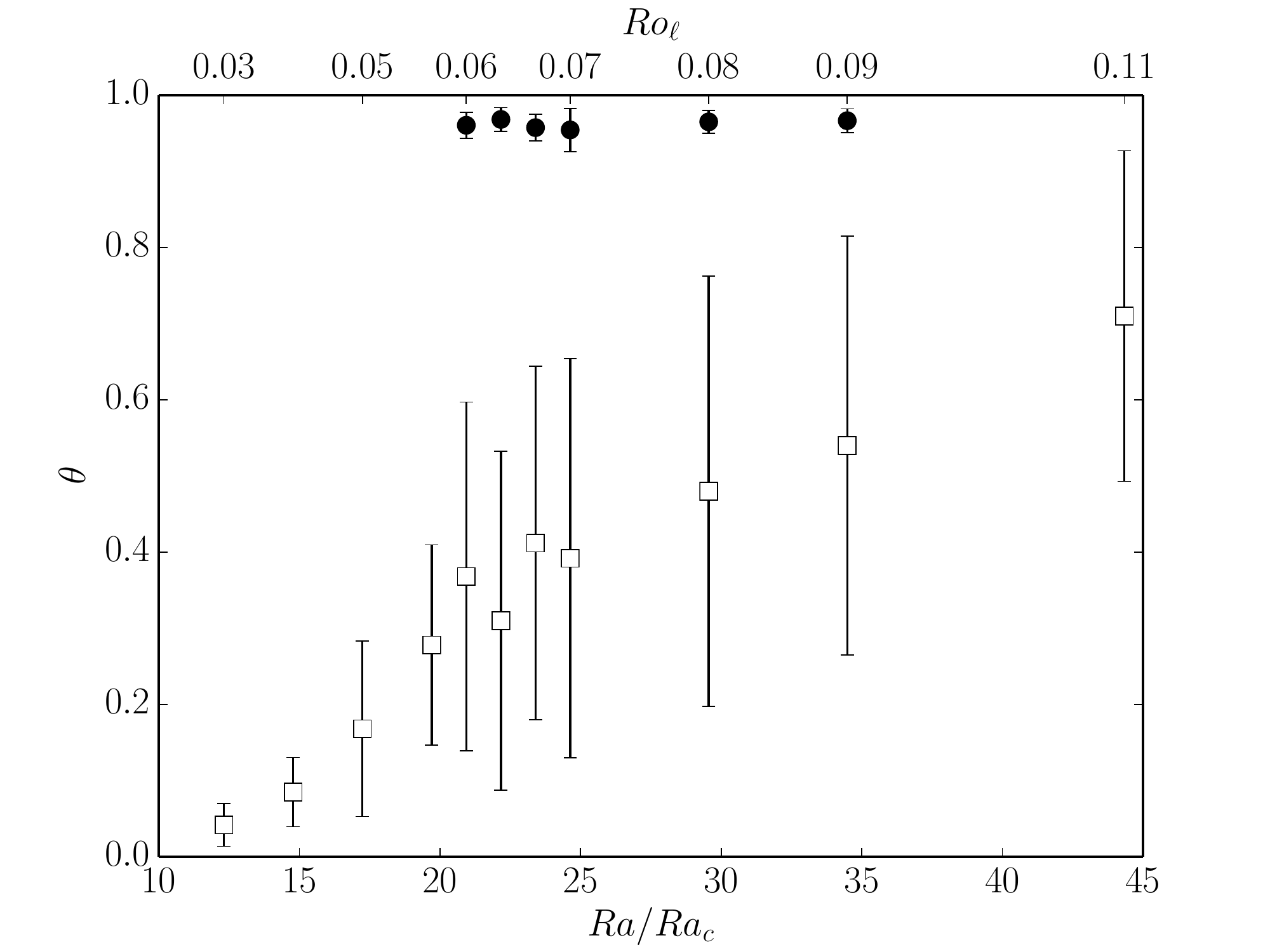} 

  \caption{(a): Evolution of the modified tilt angle $\theta$ defined
  by Eq.~\eqref{e_theta} for multipolar dynamos as a function of
  $\Ra/\Rac$ and \Pm. Colour scale ranges from white ($\theta=0$) to
  black ($\theta=0.7$). (b): $\theta$ as a function of $\Ra/\Rac$ for
  $\Pm=1$.  Upper $x$ axis corresponds to the values of \Rol{} for the
  multipolar branch. The meaning of the symbol shapes is defined in
  the caption of Fig.~\ref{f_RA_PM_01}.  Error bars represent the
  standard deviations.  }\label{f_RA_PM_theta_01}
\end{figure}

To further characterize the emergence of multipolar dynamos with a
significant equatorial dipole contribution, we plot the values of the
modified tilt angle $\theta$ in the parameter space in
Fig.~\ref{f_RA_PM_theta_01}(a).
Low values of $\theta$ are characteristic of an equatorial dipole on
the surface of the outer sphere and they appear to be preferably
localized close to the dynamo threshold of the multipolar branch, at
low Rayleigh and magnetic Prandtl numbers. In our case, dynamos with a
stronger equatorial dipole component belong to the class of
multipolar dynamos, but since they are always close to the threshold,
fewer modes are likely to be excited. As the Rayleigh number or the
magnetic Prandtl number is increased, the dipole axis is not stable
anymore but fluctuates in the interval $[0,\pi]$, which is typical of
polarity reversals for multipolar dynamos \citep{kutzner02}.  This
evolution is illustrated in Fig.~\ref{f_RA_PM_theta_01}(b) for dynamos
at $\Pm=1$.  For this subset of models, we computed the percentage of
the non-axisymmetric magnetic energy density with respect to the total
magnetic energy density $\Em$ and saw that it tends to increase from
85\% on average for multipolar dynamos up to 93\% as the Rayleigh
number is decreased.

\begin{figure}[htbp]  
 \centering
    (a)
    \includegraphics[width=\figsl, clip=true, trim=0 0
      0 0]{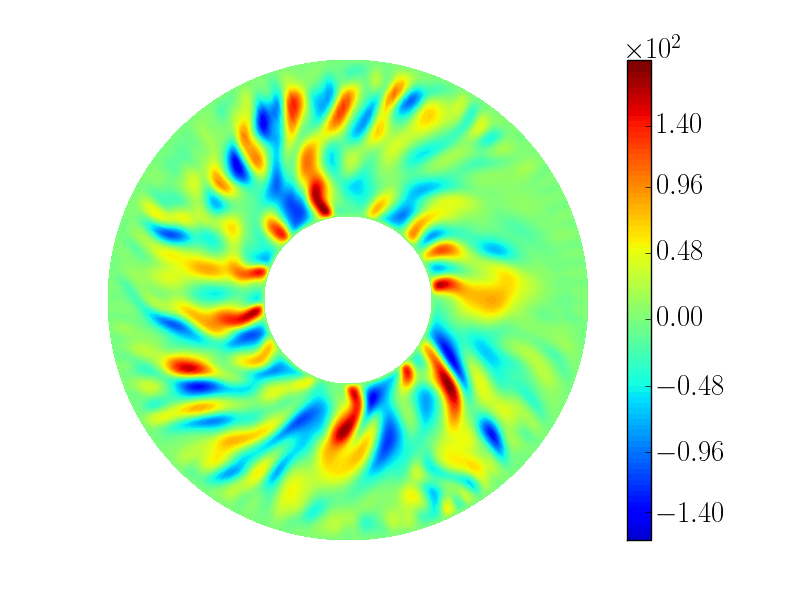} 

   (b)
    \includegraphics[width=\figsl, clip=true, trim=0 0
      0 0]{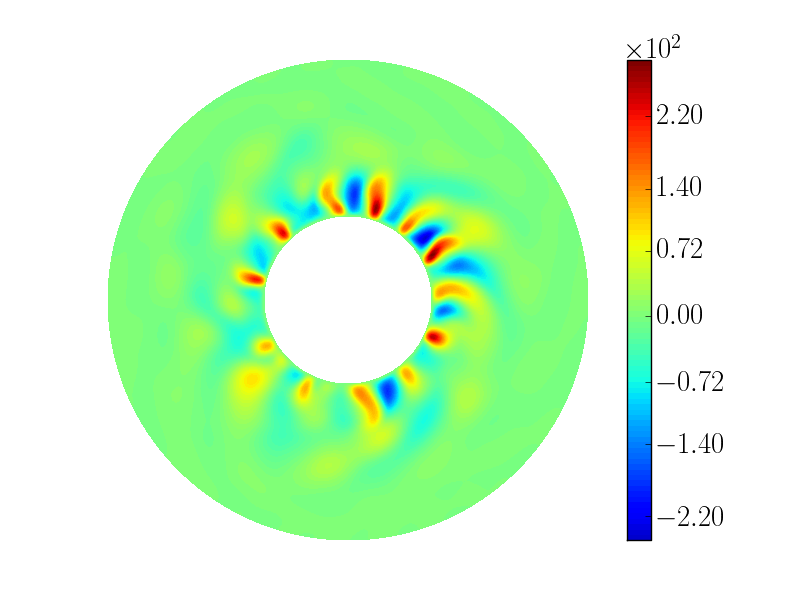}

    \caption{Equatorial cross section of the radial component of the
      velocity, with $E=10^{-4}$, $\Prandtl=1$. (a): $g \propto r\,$ and
      $\Ra/\Rac=9.0\,$, $Pm=1$.  (b): $g \propto 1/r^2$ and
      $\Ra/\Rac=9.9\,$, $Pm=1.2$.  }\label{f_vr}
\end{figure}

\begin{figure}[htbp]  
 \centering
    (a)
    \includegraphics[width=\figsl, clip=true, trim=0 0
      0 0]{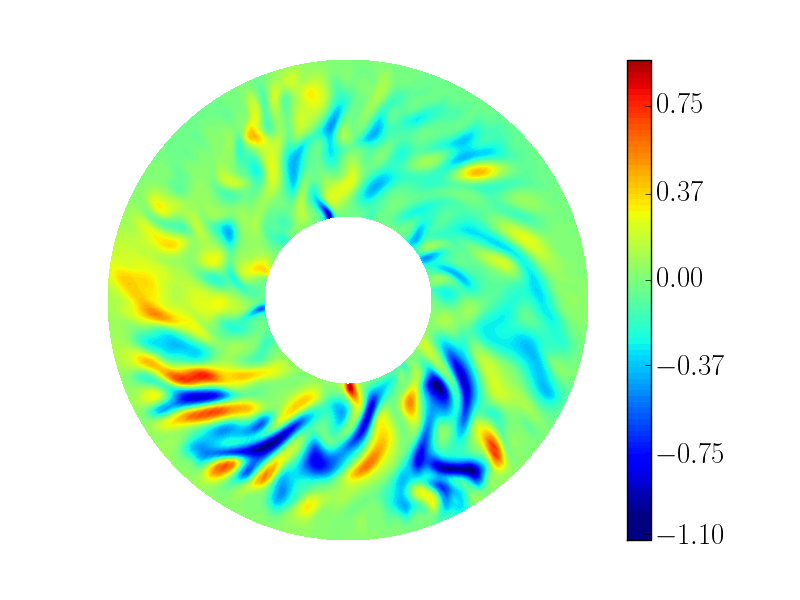}         

    (b)
    \includegraphics[width=\figsl, clip=true, trim=0 0
          0 0]{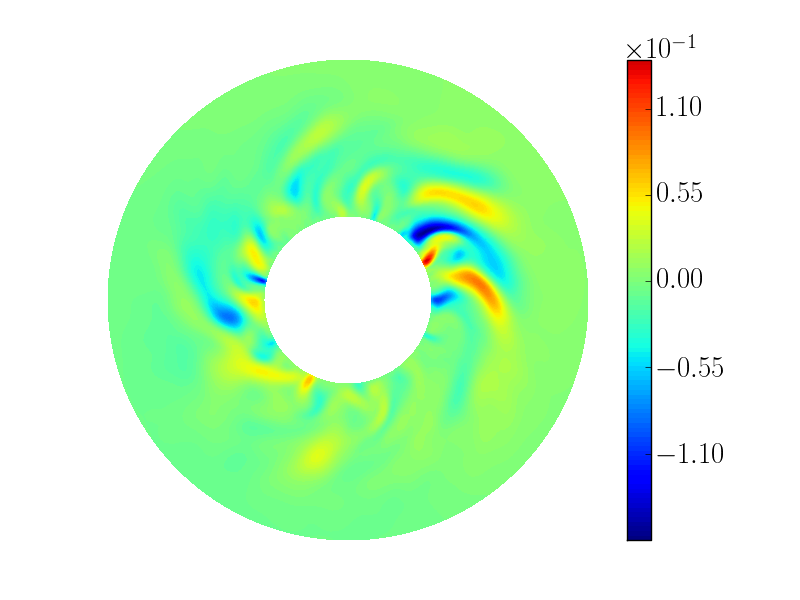} 

    (c)
    \includegraphics[width=\figsl, clip=true, trim=0 0
          0 0]{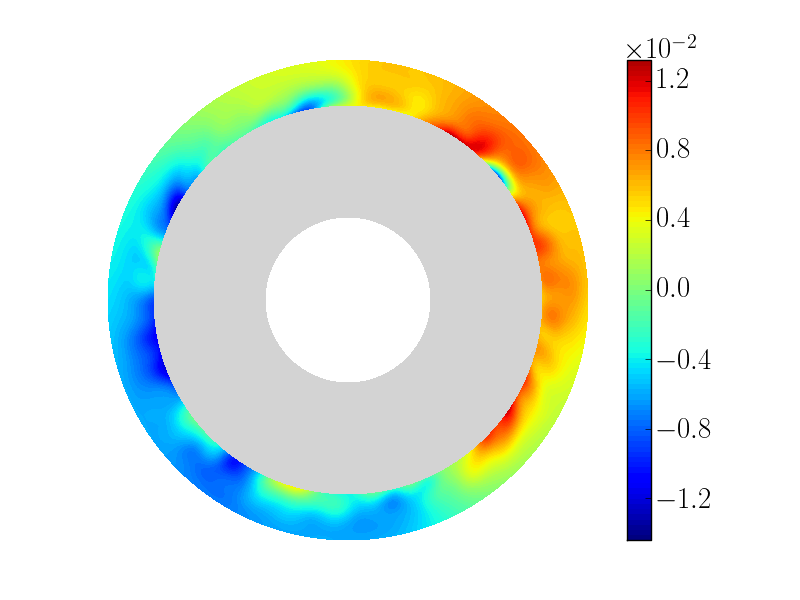}

    \caption{Equatorial cross section of the radial component of the
      magnetic field,with $E=10^{-4}$, $\Prandtl=1$. (a): $g \propto
      r\,$ and $\Ra/\Rac = 9.0\,$, $Pm=1$. (b),(c): $g \propto 1/r^2$
      and $\Ra/\Rac=9.9\,$, $Pm=1.2$. Colour in figure~(c) has been
      rescaled to highlight the emergence of a $m=1$ mode at the outer
      sphere.  }\label{f_br}
\end{figure}

Part of the changes reported in \cite{schrinner2014} about
anelastic dynamos simulations do not seem to come from the
stratified reference density profile, but from the choice of a gravity
profile proportional to $1/r^2$. This profile differs from the gravity
profile proportional to $r$ that was used for Boussinesq simulations
and is actually the only significant difference between previous
studies and our low $\Nrho$ simulations. As a consequence, convection
cells form and stay closer to the inner sphere, as we can see in
Fig.~\ref{f_vr}. We compare here equatorial cuts of the radial
component of the velocity for both choices of gravity profile.
This strong difference in the flow reflects on the localization of the
active dynamo regions, as we can see in the corresponding cuts of the
radial component of the magnetic field in Fig.~\ref{f_br}.
With a gravity profile proportional to $1/r^2$, the magnetic field is
mainly generated close to the inner sphere where the convection cells
form. Consequently, our measure of the dipole field strength \fdip{}
at the surface of the outer sphere appears to be biased, since it will
essentially be sensitive to the less diffusive large scale modes. This
filter effect is likely to be responsible for the increase in \fdip{}
we reported in some anelastic dynamo models. However, for higher
density stratification $\Nrho=3$ and Prandtl numbers $\Prandtl=2$ and
$\Pm=4$, \cite{schrinner2014} identify equatorial dipole dynamos with
a $m=1$ component that is not localized on the outer sphere and for
which the present mechanism will not be relevant.

\section{Conclusion}

In this paper, we focussed on very weakly stratified anelastic dynamo
models with a central mass distribution. We investigated the
bifurcations between the dipolar and multipolar dynamo branches and
recovered in parts the behaviour that has been observed for Boussinesq
models with a uniform mass distribution. In addition, we show that the
dipolar branch can now lose its stability and switch to the multipolar
branch at low Rayleigh and magnetic Prandtl numbers. The multipolar
dynamos that are observed in this restricted parameter regime usually
present a stronger equatorial dipole component at the surface of the
outer sphere. When increasing the Rayleigh number at higher \Pm{}, it
seems that the mean zonal flow does not grow fast enough to maintain
the multipolar solution, and we identified several cases where the
multipolar branch switches to the dipolar solution.

This study has shed interesting light on the recent systematic
parameter study of spherical anelastic dynamo models started
by \cite{schrinner2014}. Focussing on weakly stratified dynamo models,
we showed that magnetic field configurations with a significant
equatorial dipole contribution can already be observed in the
Boussinesq limit.  In the parameter regime we investigate, our study
reveals that the choice of gravity profile has a strong influence on
the fluid flow and thus on the dynamo generated magnetic field,
depending whether one considers a uniform or a central mass
distribution.  In the parameter space, we showed that multipolar
dynamos with a significant equatorial dipole contribution are
preferably observed close to the dynamo threshold. This is reminiscent
of the results of \cite{aubert04}, who studied the competition between
axial and equatorial dipolar dynamos when varying the aspect ratio of
the spherical shell in geodynamo models.

However, the filter effect we highlight here focusses on the topology
of the magnetic field at the outer surface of the models. As a
consequence, it will not be able to explain the stronger equatorial
dipole component for all models in \cite{schrinner2014}.
Independently, \cite{cole2014} also report the discovery of an
azimuthal dynamo wave of a $m=1$ mode in numerical simulations
corresponding to higher density stratification. Their Coriolis number
plays a similar role to the inverse of our local Rossby number.  Upper
$x$ axis of Fig.~\ref{f_RA_PM_theta_01}(b) shows that equatorial
dipole configurations are favoured with the decrease in \Rol{}.

Observational results from photometry \citep{hackman2013} and
spectropolarimetry \citep{kochukov2013} of rapidly rotating cool
active stars reveal that the surface magnetic field of these objects
can be highly non-axisymmetric. Further investigation of direct
numerical simulations is therefore required to better understand the
influence of the Prandtl number and the density stratification on the
magnetic field topology.
\begin{acknowledgements}
This work was granted access to the HPC resources of MesoPSL financed
by the Region Ile de France and the project Equip@Meso (reference
ANR-10-EQPX-29-01) of the programme Investissements d'Avenir
supervised by the Agence Nationale pour la Recherche. Numerical
simulations were also carried out at CEMAG and TGCC computing centres
(GENCI project x2013046698). L. P. acknowledges financial support from
``Programme National de Physique Stellaire'' (PNPS) of CNRS/INSU,
France.

\end{acknowledgements}

\bibliographystyle{aa}
\bibliography{schrinner,raph}
\appendix

\section{Scaling laws}
\label{s_scaling}

For our samples of models we compute  the usual scaling laws that have
been derived for the magnetic and velocity fields and the convective
heat flux
\citep{christensen06,yadav12,stelzer13,yadav13}. As in
\cite{schrinner2014}, we do not attempt to solve any secondary
dependence on $\Pm$ because we do not vary this parameter on a wide
enough range. We transform our problem to a linear one by taking the
logarithm and look for a law of the form $\ln \hat{y} = \beta + \alpha
\ln x$.  To quantify the misfit between data and fitted values, we
follow \cite{christensen06} and compute the relative misfit
\begin{equation}
  \chir = \sqrt{\frac{1}{n} \sum_{i=1}^n \left(
    \frac{y_i-\hat{y_i}}{y_i} \right)^2}
  \,,
\end{equation}
where $\hat{y_i}$ stands for predicted values, $y_i$ for measured
values, and $n$ is the number of data.  Our results are summarized in
Table~\ref{t_coef} and compared with those found in
\cite{schrinner2014} and \cite{yadav12} in Table~\ref{t_comp}. 
We did not find significant differences between the anelastic scalings
and the scaling we obtained in the Boussinesq limit with the same mass
distribution. The coefficients we obtained seem closer on average to
the coefficients obtained by \cite{yadav12} with Boussinesq models
with a uniform mass distribution. However, it is not possible to
deduce from our data set any influence of \Nrho{} on the different
coefficients of the scaling laws. In our models, the Nusselt number
evaluated at the surface of the inner sphere $S_{\!i}$ is defined by
\begin{equation}
  Nu_\mathrm{bot} = -\frac{(\exp{(N_\varrho)}-1)w_ir_i^2}{4\pi
    nc_1}\int_{S_{\!i}}\frac{\partial s}{\partial
    r}\sin\theta \dif \theta \dif \phi \,, \label{nu_bot}
\end{equation}
which enables a flux based Rayleigh number to be defined,
\begin{equation}
Ra_Q=Nu^\star\,\frac{Ra\, E^2}{\ro^2\, Pr} 
\quad \text{with} \quad 
Nu^\star = (Nu_\mathrm{bot}-1)\,\frac{E}{Pr} 
\,, \label{def_raq}
\end{equation}
which is used in derivating scaling laws, together with the
fraction of ohmic to total dissipation \fohm{}.

\begin{table}[htbp]
\caption{Summary of the coefficients obtained for the different
  scaling laws, with their standard error from the linear
  regression.}\label{t_coef}
\begin{tabular}{ccccc}
\hline\hline
Scaling   & $\alpha$   & $\beta$  & \chir\\
\hline
\Lof \\
multipolar & $0.29 \pm 0.01 $&$ -0.52 \pm 0.1 $ &$0.06$ \\
dipolar    & $0.32 \pm 0.02 $&$  -0.08 \pm 0.2 $ &$0.06$ \\[2mm] 
\Ro        &$ 0.39 \pm 0.01 $&$  0.35 \pm 0.2 $ &$0.1$\\[2mm]  
$Nu^\star$ &$ 0.58 \pm 0.003 $&$  -1.16 \pm 0.04 $       &$0.03$\\ 
\hline
\end{tabular}
\end{table}

\begin{table}[htbp]
\caption{Comparison between the different scaling laws obtained with
  different dynamo models. RPD refers to the present study, SPRD to
  \cite{schrinner2014}, and YGC to \cite{yadav12},
  respectively.}
\label{t_comp}
\centering
\begin{tabular}{lllllll}
\hline \hline
\multicolumn{1}{c}{Scaling}&\multicolumn{2}{c}{RPD}&\multicolumn{2}{c}{SPRD}&\multicolumn{2}{c}{YGC}\\
& \multicolumn{1}{c}{c} & \multicolumn{1}{c}{x}
& \multicolumn{1}{c}{c} & \multicolumn{1}{c}{x} &\multicolumn{1}{c}{c}
& \multicolumn{1}{c}{x} \\ 
\hline
\multicolumn{1}{c}{$g$}&\multicolumn{2}{c}{$1/r^2$}&\multicolumn{2}{c}{$1/r^2$}&\multicolumn{2}{c}{$r$}\\
\hline
\multicolumn{1}{c}{$\Nrho$}&\multicolumn{2}{c}{$\leq 0.1$}&\multicolumn{2}{c}{$\in \left[0.5, 4\right]$}&\multicolumn{2}{c}{$0$}\\
\hline 
 $\Lof=c\,Ra_Q^x$ \\
multipolar &0.59 & 0.29 & 1.19 & 0.34 & 0.65 & 0.35 \\
dipolar    &0.92 & 0.32 & 1.58 & 0.35 & 1.08 & 0.37 \\[2mm] 
$Ro=c\,Ra_Q^x$  & 1.42 & 0.39  & 1.66 & 0.42  &
1.79\tablefootmark{a} &  0.44\tablefootmark{a} \\
  &     &     & &        & 0.73\tablefootmark{b} & 0.39\tablefootmark{b}\\[2mm] 
$Nu^\star=c\,Ra_Q^{x}$ & 0.31 & 0.58 & 0.25 & 0.59  &0.06& 0.52\\
\hline
\end{tabular}
\tablefoot{\cite{yadav12} distinguished between dipolar and multipolar dynamos
  for their Rossby number scaling, whereas we derived {a single} power
  law for both classes of dynamo models.  \tablefoottext{a}
  {Multipolar models.}\tablefoottext{b} {Dipolar models.}}
\end{table}

\section{Numerical models}
\label{s_num}

\onecolumn
\input{tables/table_an_Np01}

\end{document}

%% file: tables/table_an_Np01.tex
\begin{longtable}{ccccccccccc}
\caption{Overview of the simulations carried out with $E =
  10^{-4}$, $\Prandtl=1$, $\aspectratio=0.35$, $n=2$, and $\Nrho=0.1$.}\\
\hline\hline
$ \text{Model} $&$ \Ra $&$ Pm $&$ Ro $&$ \Rol $&$ \lc $&$ Lo $&$ Nu $&$ \fdip $&$ \fdipAX $&$ \fohm $\\
\hline
\endfirsthead
\caption{(continued)}\\
\hline\hline
$ \text{Model} $ & $ \Ra $ & $ Pm $ & $ Ro $ & $ \Rol $ & $ \lc $ & $ Lo $ & $ Nu $ & $ \fdip $ & $ \fdipAX $ & $ \fohm $\\
\hline
\endhead
\hline
\endfoot
$ 1\mm $ & $ 1.50 \times 10^{6} $ & $ 2.00 $ & $ 3.6 \times 10^{-3} $ & $ 2.0 \times 10^{-2} $ & $ 1.8 \times 10^{1} $ & $ 2.1 \times 10^{-3} $ & $ 1.4 $ & $ 4.9 \times 10^{-1} $ & $ 8.0 \times 10^{-2} $ & $ 8.0 \times 10^{-2} $ \\
$ 2\mm $ & $ 1.50 \times 10^{6} $ & $ 3.00 $ & $ 3.8 \times 10^{-3} $ & $ 2.0 \times 10^{-2} $ & $ 1.6 \times 10^{1} $ & $ 3.0 \times 10^{-3} $ & $ 1.5 $ & $ 3.2 \times 10^{-1} $ & $ 9.5 \times 10^{-2} $ & $ 1.5 \times 10^{-1} $ \\
$ 3\mm $ & $ 2.00 \times 10^{6} $ & $ 1.20 $ & $ 6.8 \times 10^{-3} $ & $ 2.6 \times 10^{-2} $ & $ 1.7 \times 10^{1} $ & $ 2.5 \times 10^{-3} $ & $ 1.7 $ & $ 6.7 \times 10^{-1} $ & $ 2.7 \times 10^{-2} $ & $ 8.1 \times 10^{-2} $ \\
$ 4\mm $ & $ 2.00 \times 10^{6} $ & $ 1.50 $ & $ 6.2 \times 10^{-3} $ & $ 2.7 \times 10^{-2} $ & $ 1.7 \times 10^{1} $ & $ 3.1 \times 10^{-3} $ & $ 1.7 $ & $ 5.7 \times 10^{-1} $ & $ 4.3 \times 10^{-2} $ & $ 1.3 \times 10^{-1} $ \\
$ 5\mm $ & $ 2.00 \times 10^{6} $ & $ 2.00 $ & $ 5.1 \times 10^{-3} $ & $ 2.7 \times 10^{-2} $ & $ 1.7 \times 10^{1} $ & $ 3.6 \times 10^{-3} $ & $ 1.8 $ & $ 5.2 \times 10^{-1} $ & $ 2.5 \times 10^{-1} $ & $ 1.7 \times 10^{-1} $ \\
$ 6\mm $ & $ 2.00 \times 10^{6} $ & $ 2.75 $ & $ 5.2 \times 10^{-3} $ & $ 2.8 \times 10^{-2} $ & $ 1.7 \times 10^{1} $ & $ 4.0 \times 10^{-3} $ & $ 1.8 $ & $ 3.0 \times 10^{-1} $ & $ 8.0 \times 10^{-3} $ & $ 2.1 \times 10^{-1} $ \\
$ 7\mm $ & $ 2.00 \times 10^{6} $ & $ 2.90 $ & $ 5.2 \times 10^{-3} $ & $ 2.3 \times 10^{-2} $ & $ 1.6 \times 10^{1} $ & $ 4.1 \times 10^{-3} $ & $ 1.8 $ & $ 6.4 \times 10^{-1} $ & $ 1.3 \times 10^{-1} $ & $ 1.9 \times 10^{-1} $ \\
$ 8\mm $ & $ 2.00 \times 10^{6} $ & $ 3.00 $ & $ 5.9 \times 10^{-3} $ & $ 2.6 \times 10^{-2} $ & $ 1.6 \times 10^{1} $ & $ 4.1 \times 10^{-3} $ & $ 1.8 $ & $ 3.7 \times 10^{-1} $ & $ 2.4 \times 10^{-1} $ & $ 1.9 \times 10^{-1} $ \\
$ 9\mm $ & $ 2.00 \times 10^{6} $ & $ 4.00 $ & $ 5.6 \times 10^{-3} $ & $ 2.6 \times 10^{-2} $ & $ 1.6 \times 10^{1} $ & $ 4.5 \times 10^{-3} $ & $ 1.8 $ & $ 3.0 \times 10^{-1} $ & $ 2.1 \times 10^{-1} $ & $ 2.2 \times 10^{-1} $ \\
$ 9\dd $ & $ 2.00 \times 10^{6} $ & $ 4.00 $ & $ 5.3 \times 10^{-3} $ & $ 2.2 \times 10^{-2} $ & $ 1.4 \times 10^{1} $ & $ 6.6 \times 10^{-3} $ & $ 1.7 $ & $ 8.2 \times 10^{-1} $ & $ 8.2 \times 10^{-1} $ & $ 4.0 \times 10^{-1} $ \\
$ 10\mm $ & $ 2.00 \times 10^{6} $ & $ 5.00 $ & $ 5.5 \times 10^{-3} $ & $ 2.6 \times 10^{-2} $ & $ 1.6 \times 10^{1} $ & $ 4.7 \times 10^{-3} $ & $ 1.8 $ & $ 3.4 \times 10^{-1} $ & $ 2.5 \times 10^{-1} $ & $ 2.3 \times 10^{-1} $ \\
$ 10\dd $ & $ 2.00 \times 10^{6} $ & $ 5.00 $ & $ 5.3 \times 10^{-3} $ & $ 2.6 \times 10^{-2} $ & $ 1.4 \times 10^{1} $ & $ 6.6 \times 10^{-3} $ & $ 1.8 $ & $ 8.4 \times 10^{-1} $ & $ 8.4 \times 10^{-1} $ & $ 3.9 \times 10^{-1} $ \\
$ 11\mm $ & $ 2.00 \times 10^{6} $ & $ 6.00 $ & $ 5.4 \times 10^{-3} $ & $ 2.6 \times 10^{-2} $ & $ 1.6 \times 10^{1} $ & $ 4.8 \times 10^{-3} $ & $ 1.8 $ & $ 3.4 \times 10^{-1} $ & $ 3.1 \times 10^{-1} $ & $ 2.5 \times 10^{-1} $ \\
$ 11\dd $ & $ 2.00 \times 10^{6} $ & $ 6.00 $ & $ 5.7 \times 10^{-3} $ & $ 2.2 \times 10^{-2} $ & $ 1.3 \times 10^{1} $ & $ 9.0 \times 10^{-3} $ & $ 1.9 $ & $ 6.7 \times 10^{-1} $ & $ 6.7 \times 10^{-1} $ & $ 4.9 \times 10^{-1} $ \\
$ 12\mm $ & $ 2.50 \times 10^{6} $ & $ 1.00 $ & $ 8.0 \times 10^{-3} $ & $ 3.3 \times 10^{-2} $ & $ 1.6 \times 10^{1} $ & $ 4.1 \times 10^{-3} $ & $ 2.0 $ & $ 6.5 \times 10^{-1} $ & $ 6.0 \times 10^{-2} $ & $ 1.8 \times 10^{-1} $ \\
$ 13\mm $ & $ 3.00 \times 10^{6} $ & $ 1.00 $ & $ 7.8 \times 10^{-3} $ & $ 4.0 \times 10^{-2} $ & $ 1.6 \times 10^{1} $ & $ 5.3 \times 10^{-3} $ & $ 2.3 $ & $ 5.8 \times 10^{-1} $ & $ 1.0 \times 10^{-1} $ & $ 2.2 \times 10^{-1} $ \\
$ 14\mm $ & $ 3.00 \times 10^{6} $ & $ 2.00 $ & $ 8.9 \times 10^{-3} $ & $ 4.0 \times 10^{-2} $ & $ 1.6 \times 10^{1} $ & $ 6.4 \times 10^{-3} $ & $ 2.4 $ & $ 3.6 \times 10^{-1} $ & $ 2.1 \times 10^{-1} $ & $ 2.6 \times 10^{-1} $ \\
$ 14\dd $ & $ 3.00 \times 10^{6} $ & $ 2.00 $ & $ 7.4 \times 10^{-3} $ & $ 3.7 \times 10^{-2} $ & $ 1.5 \times 10^{1} $ & $ 8.1 \times 10^{-3} $ & $ 2.4 $ & $ 8.3 \times 10^{-1} $ & $ 7.9 \times 10^{-1} $ & $ 3.9 \times 10^{-1} $ \\
$ 15\mm $ & $ 3.00 \times 10^{6} $ & $ 3.00 $ & $ 8.7 \times 10^{-3} $ & $ 4.0 \times 10^{-2} $ & $ 1.6 \times 10^{1} $ & $ 6.8 \times 10^{-3} $ & $ 2.5 $ & $ 2.9 \times 10^{-1} $ & $ 1.3 \times 10^{-1} $ & $ 2.8 \times 10^{-1} $ \\
$ 15\dd $ & $ 3.00 \times 10^{6} $ & $ 3.00 $ & $ 7.5 \times 10^{-3} $ & $ 3.7 \times 10^{-2} $ & $ 1.6 \times 10^{1} $ & $ 8.4 \times 10^{-3} $ & $ 2.3 $ & $ 7.8 \times 10^{-1} $ & $ 7.7 \times 10^{-1} $ & $ 4.0 \times 10^{-1} $ \\
$ 16\mm $ & $ 3.00 \times 10^{6} $ & $ 4.00 $ & $ 8.2 \times 10^{-3} $ & $ 3.9 \times 10^{-2} $ & $ 1.6 \times 10^{1} $ & $ 7.2 \times 10^{-3} $ & $ 2.4 $ & $ 2.8 \times 10^{-1} $ & $ 1.8 \times 10^{-1} $ & $ 3.1 \times 10^{-1} $ \\
$ 16\dd $ & $ 3.00 \times 10^{6} $ & $ 4.00 $ & $ 7.5 \times 10^{-3} $ & $ 3.7 \times 10^{-2} $ & $ 1.6 \times 10^{1} $ & $ 8.7 \times 10^{-3} $ & $ 2.4 $ & $ 7.3 \times 10^{-1} $ & $ 7.2 \times 10^{-1} $ & $ 4.0 \times 10^{-1} $ \\
$ 17\dd $ & $ 3.00 \times 10^{6} $ & $ 6.00 $ & $ 7.8 \times 10^{-3} $ & $ 3.7 \times 10^{-2} $ & $ 1.5 \times 10^{1} $ & $ 9.2 \times 10^{-3} $ & $ 2.5 $ & $ 7.3 \times 10^{-1} $ & $ 7.3 \times 10^{-1} $ & $ 4.0 \times 10^{-1} $ \\
$ 18\mm $ & $ 3.50 \times 10^{6} $ & $ 1.00 $ & $ 9.1 \times 10^{-3} $ & $ 4.6 \times 10^{-2} $ & $ 1.6 \times 10^{1} $ & $ 6.5 \times 10^{-3} $ & $ 2.6 $ & $ 5.0 \times 10^{-1} $ & $ 1.8 \times 10^{-1} $ & $ 2.5 \times 10^{-1} $ \\
$ 19\mm $ & $ 4.00 \times 10^{6} $ & $ 0.50 $ & $ 1.7 \times 10^{-2} $ & $ 4.8 \times 10^{-2} $ & $ 1.6 \times 10^{1} $ & $ 4.9 \times 10^{-3} $ & $ 2.7 $ & $ 1.9 \times 10^{-1} $ & $ 1.3 \times 10^{-1} $ & $ 1.9 \times 10^{-1} $ \\
$ 20\mm $ & $ 4.00 \times 10^{6} $ & $ 1.00 $ & $ 1.0 \times 10^{-2} $ & $ 5.3 \times 10^{-2} $ & $ 1.6 \times 10^{1} $ & $ 7.7 \times 10^{-3} $ & $ 3.0 $ & $ 5.0 \times 10^{-1} $ & $ 2.2 \times 10^{-1} $ & $ 3.1 \times 10^{-1} $ \\
$ 21\mm $ & $ 4.00 \times 10^{6} $ & $ 3.00 $ & $ 1.1 \times 10^{-2} $ & $ 5.2 \times 10^{-2} $ & $ 1.6 \times 10^{1} $ & $ 9.0 \times 10^{-3} $ & $ 3.1 $ & $ 3.0 \times 10^{-1} $ & $ 2.4 \times 10^{-1} $ & $ 3.4 \times 10^{-1} $ \\
$ 21\dd $ & $ 4.00 \times 10^{6} $ & $ 3.00 $ & $ 1.0 \times 10^{-2} $ & $ 4.9 \times 10^{-2} $ & $ 1.6 \times 10^{1} $ & $ 1.1 \times 10^{-2} $ & $ 3.1 $ & $ 6.5 \times 10^{-1} $ & $ 6.5 \times 10^{-1} $ & $ 4.2 \times 10^{-1} $ \\
$ 22\dd $ & $ 4.00 \times 10^{6} $ & $ 4.00 $ & $ 1.0 \times 10^{-2} $ & $ 4.9 \times 10^{-2} $ & $ 1.5 \times 10^{1} $ & $ 1.2 \times 10^{-2} $ & $ 3.1 $ & $ 6.6 \times 10^{-1} $ & $ 6.5 \times 10^{-1} $ & $ 4.3 \times 10^{-1} $ \\
$ 23\mm $ & $ 4.25 \times 10^{6} $ & $ 1.00 $ & $ 1.1 \times 10^{-2} $ & $ 5.7 \times 10^{-2} $ & $ 1.6 \times 10^{1} $ & $ 8.1 \times 10^{-3} $ & $ 3.2 $ & $ 4.0 \times 10^{-1} $ & $ 2.6 \times 10^{-1} $ & $ 2.8 \times 10^{-1} $ \\
$ 23\dd $ & $ 4.25 \times 10^{6} $ & $ 1.00 $ & $ 1.0 \times 10^{-2} $ & $ 5.1 \times 10^{-2} $ & $ 1.6 \times 10^{1} $ & $ 1.1 \times 10^{-2} $ & $ 3.2 $ & $ 9.4 \times 10^{-1} $ & $ 9.4 \times 10^{-1} $ & $ 4.4 \times 10^{-1} $ \\
$ 24\mm $ & $ 4.50 \times 10^{6} $ & $ 1.00 $ & $ 1.2 \times 10^{-2} $ & $ 6.0 \times 10^{-2} $ & $ 1.6 \times 10^{1} $ & $ 8.7 \times 10^{-3} $ & $ 3.4 $ & $ 3.4 \times 10^{-1} $ & $ 1.1 \times 10^{-1} $ & $ 3.0 \times 10^{-1} $ \\
$ 24\dd $ & $ 4.50 \times 10^{6} $ & $ 1.00 $ & $ 1.1 \times 10^{-2} $ & $ 5.5 \times 10^{-2} $ & $ 1.6 \times 10^{1} $ & $ 1.2 \times 10^{-2} $ & $ 3.4 $ & $ 9.3 \times 10^{-1} $ & $ 9.3 \times 10^{-1} $ & $ 4.4 \times 10^{-1} $ \\
$ 25\mm $ & $ 4.75 \times 10^{6} $ & $ 1.00 $ & $ 1.2 \times 10^{-2} $ & $ 6.2 \times 10^{-2} $ & $ 1.6 \times 10^{1} $ & $ 9.2 \times 10^{-3} $ & $ 3.6 $ & $ 3.5 \times 10^{-1} $ & $ 2.7 \times 10^{-1} $ & $ 3.0 \times 10^{-1} $ \\
$ 25\dd $ & $ 4.75 \times 10^{6} $ & $ 1.00 $ & $ 1.2 \times 10^{-2} $ & $ 5.7 \times 10^{-2} $ & $ 1.5 \times 10^{1} $ & $ 1.2 \times 10^{-2} $ & $ 3.6 $ & $ 9.2 \times 10^{-1} $ & $ 9.2 \times 10^{-1} $ & $ 4.5 \times 10^{-1} $ \\
$ 26\mm $ & $ 5.00 \times 10^{6} $ & $ 0.50 $ & $ 1.7 \times 10^{-2} $ & $ 6.6 \times 10^{-2} $ & $ 1.6 \times 10^{1} $ & $ 8.1 \times 10^{-3} $ & $ 3.7 $ & $ 2.9 \times 10^{-1} $ & $ 2.0 \times 10^{-1} $ & $ 2.5 \times 10^{-1} $ \\
$ 27\mm $ & $ 5.00 \times 10^{6} $ & $ 1.00 $ & $ 1.3 \times 10^{-2} $ & $ 6.6 \times 10^{-2} $ & $ 1.6 \times 10^{1} $ & $ 9.7 \times 10^{-3} $ & $ 3.8 $ & $ 3.7 \times 10^{-1} $ & $ 2.3 \times 10^{-1} $ & $ 3.2 \times 10^{-1} $ \\
$ 28\dd $ & $ 5.00 \times 10^{6} $ & $ 1.00 $ & $ 1.2 \times 10^{-2} $ & $ 6.0 \times 10^{-2} $ & $ 1.5 \times 10^{1} $ & $ 1.3 \times 10^{-2} $ & $ 3.8 $ & $ 9.2 \times 10^{-1} $ & $ 9.2 \times 10^{-1} $ & $ 4.6 \times 10^{-1} $ \\
$ 29\mm $ & $ 5.00 \times 10^{6} $ & $ 2.00 $ & $ 1.4 \times 10^{-2} $ & $ 6.7 \times 10^{-2} $ & $ 1.6 \times 10^{1} $ & $ 1.1 \times 10^{-2} $ & $ 3.9 $ & $ 2.5 \times 10^{-1} $ & $ 2.1 \times 10^{-1} $ & $ 3.6 \times 10^{-1} $ \\
$ 30\dd $ & $ 5.00 \times 10^{6} $ & $ 2.00 $ & $ 1.2 \times 10^{-2} $ & $ 6.1 \times 10^{-2} $ & $ 1.6 \times 10^{1} $ & $ 1.4 \times 10^{-2} $ & $ 3.9 $ & $ 7.2 \times 10^{-1} $ & $ 7.2 \times 10^{-1} $ & $ 4.6 \times 10^{-1} $ \\
$ 31\dd $ & $ 5.00 \times 10^{6} $ & $ 3.00 $ & $ 1.2 \times 10^{-2} $ & $ 6.1 \times 10^{-2} $ & $ 1.6 \times 10^{1} $ & $ 1.4 \times 10^{-2} $ & $ 3.9 $ & $ 6.4 \times 10^{-1} $ & $ 6.3 \times 10^{-1} $ & $ 4.6 \times 10^{-1} $ \\
$ 32\dd $ & $ 5.00 \times 10^{6} $ & $ 4.00 $ & $ 1.3 \times 10^{-2} $ & $ 6.2 \times 10^{-2} $ & $ 1.6 \times 10^{1} $ & $ 1.4 \times 10^{-2} $ & $ 4.1 $ & $ 7.3 \times 10^{-1} $ & $ 7.3 \times 10^{-1} $ & $ 4.3 \times 10^{-1} $ \\
$ 33\mm $ & $ 6.00 \times 10^{6} $ & $ 1.00 $ & $ 1.5 \times 10^{-2} $ & $ 7.8 \times 10^{-2} $ & $ 1.6 \times 10^{1} $ & $ 1.2 \times 10^{-2} $ & $ 4.5 $ & $ 3.5 \times 10^{-1} $ & $ 2.9 \times 10^{-1} $ & $ 3.4 \times 10^{-1} $ \\
$ 34\dd $ & $ 6.00 \times 10^{6} $ & $ 1.00 $ & $ 1.5 \times 10^{-2} $ & $ 7.2 \times 10^{-2} $ & $ 1.5 \times 10^{1} $ & $ 1.6 \times 10^{-2} $ & $ 4.6 $ & $ 9.1 \times 10^{-1} $ & $ 9.1 \times 10^{-1} $ & $ 4.9 \times 10^{-1} $ \\
$ 35\dd $ & $ 6.00 \times 10^{6} $ & $ 3.00 $ & $ 1.4 \times 10^{-2} $ & $ 7.1 \times 10^{-2} $ & $ 1.6 \times 10^{1} $ & $ 1.7 \times 10^{-2} $ & $ 4.7 $ & $ 6.8 \times 10^{-1} $ & $ 6.8 \times 10^{-1} $ & $ 4.8 \times 10^{-1} $ \\
$ 36\dd $ & $ 6.00 \times 10^{6} $ & $ 4.00 $ & $ 1.5 \times 10^{-2} $ & $ 7.4 \times 10^{-2} $ & $ 1.6 \times 10^{1} $ & $ 1.8 \times 10^{-2} $ & $ 4.9 $ & $ 7.2 \times 10^{-1} $ & $ 7.2 \times 10^{-1} $ & $ 4.7 \times 10^{-1} $ \\
$ 37\mm $ & $ 7.00 \times 10^{6} $ & $ 0.50 $ & $ 1.8 \times 10^{-2} $ & $ 9.2 \times 10^{-2} $ & $ 1.6 \times 10^{1} $ & $ 1.2 \times 10^{-2} $ & $ 5.2 $ & $ 2.4 \times 10^{-1} $ & $ 2.1 \times 10^{-1} $ & $ 3.7 \times 10^{-1} $ \\
$ 38\mm $ & $ 7.00 \times 10^{6} $ & $ 1.00 $ & $ 2.0 \times 10^{-2} $ & $ 9.3 \times 10^{-2} $ & $ 1.6 \times 10^{1} $ & $ 1.4 \times 10^{-2} $ & $ 5.5 $ & $ 2.9 \times 10^{-1} $ & $ 2.5 \times 10^{-1} $ & $ 3.7 \times 10^{-1} $ \\
$ 38\dd $ & $ 7.00 \times 10^{6} $ & $ 1.00 $ & $ 1.7 \times 10^{-2} $ & $ 8.0 \times 10^{-2} $ & $ 1.5 \times 10^{1} $ & $ 1.8 \times 10^{-2} $ & $ 5.2 $ & $ 8.9 \times 10^{-1} $ & $ 8.9 \times 10^{-1} $ & $ 5.2 \times 10^{-1} $ \\
$ 39\dd $ & $ 7.00 \times 10^{6} $ & $ 3.00 $ & $ 1.6 \times 10^{-2} $ & $ 7.7 \times 10^{-2} $ & $ 1.5 \times 10^{1} $ & $ 2.0 \times 10^{-2} $ & $ 5.4 $ & $ 7.2 \times 10^{-1} $ & $ 7.2 \times 10^{-1} $ & $ 5.2 \times 10^{-1} $ \\
$ 40\dd $ & $ 7.00 \times 10^{6} $ & $ 4.00 $ & $ 1.7 \times 10^{-2} $ & $ 8.0 \times 10^{-2} $ & $ 1.6 \times 10^{1} $ & $ 2.1 \times 10^{-2} $ & $ 5.4 $ & $ 8.4 \times 10^{-1} $ & $ 8.4 \times 10^{-1} $ & $ 5.0 \times 10^{-1} $ \\
$ 41\mm $ & $ 9.00 \times 10^{6} $ & $ 1.00 $ & $ 2.4 \times 10^{-2} $ & $ 1.1 \times 10^{-1} $ & $ 1.5 \times 10^{1} $ & $ 1.8 \times 10^{-2} $ & $ 6.8 $ & $ 3.5 \times 10^{-1} $ & $ 3.3 \times 10^{-1} $ & $ 4.1 \times 10^{-1} $ 
\label{t_Np01}
\end{longtable}
\tablefoot{The critical Rayleigh number for the onset of convection is
  $\Rac=2.03\times 10^5$, and the corresponding critical azimuthal
  wavenumber $m_c=8$.}

%% file: paper.bbl
\begin{thebibliography}{36}
\expandafter\ifx\csname natexlab\endcsname\relax\def\natexlab#1{#1}\fi

\bibitem[{{Alboussi{\`e}re} \& {Ricard}(2013)}]{alboussiere13}
{Alboussi{\`e}re}, T. \& {Ricard}, Y. 2013, Journal of Fluid Mechanics, 725, 1

\bibitem[{{Anufriev} {et~al.}(2005){Anufriev}, {Jones}, \&
  {Soward}}]{anufriev2005}
{Anufriev}, A.~P., {Jones}, C.~A., \& {Soward}, A.~M. 2005, Physics of the
  Earth and Planetary Interiors, 152, 163

\bibitem[{{Aubert} \& {Wicht}(2004)}]{aubert04}
{Aubert}, J. \& {Wicht}, J. 2004, Earth and Planetary Science Letters, 221, 409

\bibitem[{{Berkoff} {et~al.}(2010){Berkoff}, {Kersale}, \&
  {Tobias}}]{berkoff2010}
{Berkoff}, N.~A., {Kersale}, E., \& {Tobias}, S.~M. 2010, Geophysical and
  Astrophysical Fluid Dynamics, 104, 545

\bibitem[{Boussinesq(1903)}]{boussinesq1903}
Boussinesq, J. 1903, Th{\'e}orie analytique de la chaleur, Vol.~2
  (Gauthier-Villars)

\bibitem[{{Braginsky} \& {Roberts}(1995)}]{braginsky95}
{Braginsky}, S.~I. \& {Roberts}, P.~H. 1995, Geophysical and Astrophysical
  Fluid Dynamics, 79, 1

\bibitem[{{Brown} {et~al.}(2012){Brown}, {Vasil}, \& {Zweibel}}]{brown12}
{Brown}, B.~P., {Vasil}, G.~M., \& {Zweibel}, E.~G. 2012, \apj, 756, 109

\bibitem[{{Busse} \& {Simitev}(2006)}]{busse06}
{Busse}, F.~H. \& {Simitev}, R.~D. 2006, Geophys. Astrophys. Fluid Dyn., 100,
  341

\bibitem[{{Christensen} \& {Aubert}(2006)}]{christensen06}
{Christensen}, U.~R. \& {Aubert}, J. 2006, Geophy. J. Int., 166, 97

\bibitem[{{Christensen} {et~al.}(2001){Christensen}, {Aubert}, {Cardin},
  {Dormy}, {Gibbons}, {Glatzmaier}, {Grote}, {Honkura}, {Jones}, {Kono},
  {Matsushima}, {Sakuraba}, {Takahashi}, {Tilgner}, {Wicht}, \&
  {Zhang}}]{christensen01}
{Christensen}, U.~R., {Aubert}, J., {Cardin}, P., {et~al.} 2001, Phys. Earth
  Planet. Inter., 128, 25

\bibitem[{{Cole} {et~al.}(2014){Cole}, {K{\"a}pyl{\"a}}, {Mantere}, \&
  {Brandenburg}}]{cole2014}
{Cole}, E., {K{\"a}pyl{\"a}}, P.~J., {Mantere}, M.~J., \& {Brandenburg}, A.
  2014, \apjl, 780, L22

\bibitem[{{Donati} \& {Landstreet}(2009)}]{donati09}
{Donati}, J.-F. \& {Landstreet}, J.~D. 2009, \araa, 47, 333

\bibitem[{{Dormy} {et~al.}(1998){Dormy}, {Cardin}, \& {Jault}}]{dormy98}
{Dormy}, E., {Cardin}, P., \& {Jault}, D. 1998, Earth Planet. Sci. Lett., 160,
  15

\bibitem[{{{Dormy}, E. and {Soward}, A.~M.}(2007)}]{book_dormy}
{{Dormy}, E. and {Soward}, A.~M.}, ed. 2007, {Mathematical aspects of natural
  dynamos} (CRC Press)

\bibitem[{{Duarte} {et~al.}(2013){Duarte}, {Gastine}, \& {Wicht}}]{duarte2013}
{Duarte}, L.~D.~V., {Gastine}, T., \& {Wicht}, J. 2013, Physics of the Earth
  and Planetary Interiors, 222, 22

\bibitem[{{Gastine} {et~al.}(2012){Gastine}, {Duarte}, \& {Wicht}}]{gastine12}
{Gastine}, T., {Duarte}, L., \& {Wicht}, J. 2012, \aap, 546, A19

\bibitem[{{Gilman} \& {Glatzmaier}(1981)}]{gilman81}
{Gilman}, P.~A. \& {Glatzmaier}, G.~A. 1981, \apjs, 45, 335

\bibitem[{{Gissinger} {et~al.}(2012){Gissinger}, {Petitdemange}, {Schrinner},
  \& {Dormy}}]{gissinger12}
{Gissinger}, C., {Petitdemange}, L., {Schrinner}, M., \& {Dormy}, E. 2012,
  Physical Review Letters, 108, 234501

\bibitem[{{Gough}(1969)}]{gough69}
{Gough}, D.~O. 1969, Journal of Atmospheric Sciences, 26, 448

\bibitem[{{Hackman} {et~al.}(2013){Hackman}, {Pelt}, {Mantere}, {Jetsu},
  {Korhonen}, {Granzer}, {Kajatkari}, {Lehtinen}, \&
  {Strassmeier}}]{hackman2013}
{Hackman}, T., {Pelt}, J., {Mantere}, M.~J., {et~al.} 2013, \aap, 553, A40

\bibitem[{{Jones}(2011)}]{jones2011a}
{Jones}, C.~A. 2011, Annual Review of Fluid Mechanics, 43, 583

\bibitem[{{Jones} {et~al.}(2011){Jones}, {Boronski}, {Brun}, {Glatzmaier},
  {Gastine}, {Miesch}, \& {Wicht}}]{jones11}
{Jones}, C.~A., {Boronski}, P., {Brun}, A.~S., {et~al.} 2011, \icarus, 216, 120

\bibitem[{{Kochukhov} {et~al.}(2013){Kochukhov}, {Mantere}, {Hackman}, \&
  {Ilyin}}]{kochukov2013}
{Kochukhov}, O., {Mantere}, M.~J., {Hackman}, T., \& {Ilyin}, I. 2013, \aap,
  550, A84

\bibitem[{{Kutzner} \& {Christensen}(2002)}]{kutzner02}
{Kutzner}, C. \& {Christensen}, U.~R. 2002, Physics of the Earth and Planetary
  Interiors, 131, 29

\bibitem[{{Lantz} \& {Fan}(1999)}]{lantz99}
{Lantz}, S.~R. \& {Fan}, Y. 1999, \apjs, 121, 247

\bibitem[{{Morin} {et~al.}(2010){Morin}, {Donati}, {Petit}, {Delfosse},
  {Forveille}, \& {Jardine}}]{morinD2010}
{Morin}, J., {Donati}, J.-F., {Petit}, P., {et~al.} 2010, \mnras, 407, 2269

\bibitem[{{Morin} {et~al.}(2011){Morin}, {Dormy}, {Schrinner}, \&
  {Donati}}]{morin11}
{Morin}, J., {Dormy}, E., {Schrinner}, M., \& {Donati}, J.-F. 2011, \mnras,
  418, L133

\bibitem[{{Ogura} \& {Phillips}(1962)}]{ogura62}
{Ogura}, Y. \& {Phillips}, N.~A. 1962, Journal of Atmospheric Sciences, 19, 173

\bibitem[{{Olson} \& {Christensen}(2006)}]{olson06}
{Olson}, P. \& {Christensen}, U.~R. 2006, Earth and Planetary Science Letters,
  250, 561

\bibitem[{{Reinhold} {et~al.}(2013){Reinhold}, {Reiners}, \&
  {Basri}}]{reinhold2013}
{Reinhold}, T., {Reiners}, A., \& {Basri}, G. 2013, \aap, 560, A4

\bibitem[{{Sasaki} {et~al.}(2011){Sasaki}, {Takehiro}, {Kuramoto}, \&
  {Hayashi}}]{sasaki2011}
{Sasaki}, Y., {Takehiro}, S.-i., {Kuramoto}, K., \& {Hayashi}, Y.-Y. 2011,
  Physics of the Earth and Planetary Interiors, 188, 203

\bibitem[{{Schrinner} {et~al.}(2012){Schrinner}, {Petitdemange}, \&
  {Dormy}}]{schrinner12}
{Schrinner}, M., {Petitdemange}, L., \& {Dormy}, E. 2012, \apj, 752, 121

\bibitem[{{Schrinner} {et~al.}(2014){Schrinner}, {Petitdemange}, {Raynaud}, \&
  {Dormy}}]{schrinner2014}
{Schrinner}, M., {Petitdemange}, L., {Raynaud}, R., \& {Dormy}, E. 2014, \aap,
  564, A78

\bibitem[{{Stelzer} \& {Jackson}(2013)}]{stelzer13}
{Stelzer}, Z. \& {Jackson}, A. 2013, Geophysical Journal International, 193,
  1265

\bibitem[{{Yadav} {et~al.}(2013{\natexlab{a}}){Yadav}, {Gastine}, \&
  {Christensen}}]{yadav12}
{Yadav}, R.~K., {Gastine}, T., \& {Christensen}, U.~R. 2013{\natexlab{a}},
  \icarus, 225, 185

\bibitem[{{Yadav} {et~al.}(2013{\natexlab{b}}){Yadav}, {Gastine},
  {Christensen}, \& {Duarte}}]{yadav13}
{Yadav}, R.~K., {Gastine}, T., {Christensen}, U.~R., \& {Duarte}, L.~D.~V.
  2013{\natexlab{b}}, ArXiv e-prints

\end{thebibliography}
